\documentclass[usegraphicx,usenatbib]{mn2e}

\usepackage{times}
\usepackage{graphicx,epsf}

\newcommand{\apj}{ApJ}           % Astrophysical Journal
\newcommand{\apjs}{ApJS}
\newcommand{\mnras}{MNRAS}       % Monthly Notices of the RAS
\newcommand{\aj}{AJ}
\newcommand{\araa}{ARAA}
\newcommand{\aap}{A\&A}
\newcommand{\aaps}{A\&AS}
\newcommand{\plotone}{\includegraphics[width=\columnwidth,clip=true]}

\title{Fourier Photometric Analysis of Isolated Galaxies in the context of the AMIGA project}

\author[Durbala et al.]{A. Durbala$^1$\thanks{E-mail: adriana.durbala@ua.edu}, R. Buta$^1$, J.~W.
Sulentic$^1$, and L. Verdes-Montenegro$^2$\\
$^1$Department of Physics and Astronomy, University of Alabama,
Box 870324, Tuscaloosa, AL 35487-0324, USA\\
$^2$Instituto de Astrof\'{\i}sica de Andaluc\'{\i}a, CSIC, Apdo.
3004, 18080 Granada, Spain\\}

\begin{document}

\maketitle

\begin{abstract}

We present here the results of a Fourier photometric decomposition
of a representative sample of $\sim$ 100 isolated CIG galaxies
(Catalog of Isolated Galaxies; \citealt{karachentseva73}) in the
morphological range Sb-Sc. This study is an integral part of the
AMIGA project. It complements the photometric analysis presented in
our previous paper \citep{durbala08} for the same sample of disk
galaxies by allowing a description of the spiral structure
morphology. We also estimate dynamical measures like torque strength
for bar and spiral, and also the total nonaxisymmetric torque by
assuming a constant M/L ratio, and explore the interplay between the
spiral and bar components of galaxies. Both the length (l$_{bar}$)
and the contrast (e.g. A$_{2b}$) of the Fourier bars decrease along
the morphological sequence Sb-Sbc-Sc, with bars in earlier types
being longer and showing higher contrast. The bars of Sb galaxies
are $\sim$ 3$\times$ longer than the bars in Sc types, consistent
with our previous study \citep{durbala08}. We find that the longer
bars are not necessarily stronger (as quantified by the torque
Q$_{b}$ measure), but longer bars show a higher contrast A$_{2b}$,
in very good agreement with theoretical predictions. Our data
suggests that bar and spiral components are rather independent in
the sense that the torque strengths of the two components are not
correlated. The total strength Q$_{g}$ is a very reliable tracer of
the bar strength measure Q$_{b}$, the two quantities showing a very
tight linear correlation. Comparison with a similar sample of disk
galaxies (same morphological range) extracted from the OSUBGS
\citep[Ohio State University Bright Galaxy Survey;][]{eskridge02}
survey indicates that the isolated CIG/AMIGA galaxies host
significantly longer Fourier bars and possibly show a different
distribution of spiral torque Q$_{s}$. The Fourier analysis also
revealed a potential case of counterwinding spiral structure (KIG
652/NGC 5768), which deserves further kinematic study. We find that
m = 2 (i.e. dominating two-armed pattern) is the most common spiral
arm multiplicity among the sample of Sb-Sc CIG/AMIGA galaxies
($\sim$ 40\%), m = 2 \& 3 and m = 1 \& 2 are found in $\sim$ 28\%
and $\sim$ 13\% of isolated galaxies, respectively.

\end{abstract}

\begin{keywords}
galaxies: fundamental parameters; galaxies: photometry; galaxies:
structure; galaxies: evolution; galaxies: spiral; galaxies: general
\end{keywords}

\section{Introduction}

This is our second study dedicated to a detailed photometric
characterization of isolated galaxies in the context of the AMIGA
(\textbf{A}nalysis of the interstellar \textbf{M}edium of
\textbf{I}solated \textbf{GA}laxies)
project\footnote{http://www.iaa.es/AMIGA.html}. Our first paper
\citet{durbala08} presented a dual approach to characterize the
properties of a representative sample of n $\sim$ 100 Sb-Sc isolated
galaxies: bulge/disk/bar decomposition and CAS
(Concentration/Asymmetry/Clumpiness) parametrization. The main goal
was to explore morphological type differences using quantitative
structural (photometric) analysis. In that context we quantified
structural properties of galaxies thought to be least influenced by
environment ($\sim$ zero nurture). Since one expects that
environment almost certainly increases ``dispersion'' in virtually
all galaxy measures, we wanted to constrain the best estimates of
``intrinsic dispersion'' ($\sim$ pure nature).

So far we find: i) extreme bias towards spirals (few E+S0;
\citealt{sulentic06}), ii) bias to intermediate-late type spirals,
with a clear dominance of Sb-Sc morphological types
\citep{sulentic06}, iii) the majority of CIG/AMIGA disk galaxies
host pseudobulges \citep{durbala08}, iv) the core sample of
CIG/AMIGA isolated galaxies (Sb-Sc types) tends to host larger bars
and shows lower concentration and asymmetry measures than galaxy
samples of similar morphological classification selected without
isolation criteria \citep{durbala08}.

However, neither approach in \citet{durbala08} was sensitive to the
spiral arm morphology, which is intimately connected to the global
galactic morphology. This paper presents a 2D Fourier
decomposition/analysis of the same sample explored in that previous
paper. The present study offers a complementary description not only
by incorporating the structural properties of the spiral arms, but
also allowing for dynamical measures (i.e. gravitational torque) for
bars, spiral arms and total (bar+spiral) nonaxisymmetric components.
We emphasize that these dynamical measures (see \S\S\S~3.1.2) will
be referred herein as ``strength''. There are studies
\citep[e.g.][]{athanassoula03} where other kind of measures defined
in terms of relative Fourier amplitudes are called ``strength''.
Such parameters are similar to what would be referred in our context
as ``contrast'' (see \S~4).

In the context of the AMIGA project a similar Fourier decomposition
technique was employed by \citet{verley07b} for a different sample
of isolated galaxies spanning the full range of morphological types
later than S0/a. That study explored the dynamical influence of bars
on star formation properties.

The representative collection of isolated Sb-Sc CIG/AMIGA galaxies
we have examined in the present paper (and also in
\citealt{durbala08}) constitutes a valuable control sample to test
the predictions of theoretical models regarding the coevolution and
the interplay between various galactic components. The goal is to
compare our results of the Fourier analysis for our sample of
isolated galaxies with measures from samples selected without
isolation criteria. The main question is whether we could reveal the
environmental influence on the morphology and dynamics of spiral
galaxies. We would also like to present a census of the spiral
pattern morphology, i.e. we evaluate the frequency of occurrence of
one-, two- or three-armed pattern morphology amongst our sample.

This paper is organized as follows: \S~2 presents our sample, \S~3
offers a detailed description of the data reduction and the Fourier
decomposition, \S~4 is dedicated to the analysis of the parameters
provided by Fourier decomposition. \S~5 discusses the results of
this study and \S~6 outlines the most important conclusions.
Throughout the paper we use H$_{o}$ = 75 km s$^{-1}$ Mpc$^{-1}$.

\section{Sample}

Our isolated galaxy sample is drawn from the Catalog of Isolated
Galaxies \citep[CIG;][]{karachentseva73}. We focus on Sb-Sc
morphological type, since they represent the bulk (63\%) of all
isolated AMIGA galaxies \citep{sulentic06}. The sample selection was
described in detail in \citet{durbala08}, where we studied galaxies
that have inclinations less than $\sim$ 70$^\circ$ and have i-band
images available in the Sloan Digital Sky Survey (SDSS-DR6). In our
present study we exclude one galaxy (KIG 907) because we cannot get
reliable Fourier measures. Therefore, the statistical analysis
herein will focus on a sample of n=93 galaxies.

\section{Data Reduction}

The SDSS i-band frames we use are flat-field, bias, cosmic ray and
pixel-defect corrected \citep{stoughton02}. Foreground stars were
removed from the images using IRAF task IMEDIT. Sky fitting and
subtraction were accomplished using IRAF task IMSURFIT. The
\emph{aa}, \emph{kk} and \emph{airmass} coefficients (zeropoint,
extinction coefficient and airmass) from the SDSS TsField files were
used to perform the photometric
calibration\footnote{http://www.sdss.org/dr6/algorithms/fluxcal.html}.
The surface brightness zeropoint was calculated using the following
formula:
$2.5\times\log(exptime\times0.396^{2})-2.5\times0.4\times(aa+kk\times
airmass)$, where an exposure time \emph{exptime} of 53.907 seconds
and a pixel size of 0\farcs396\ were used.

\subsection{Fourier Decomposition}

The observed light distribution in a deprojected galaxy image can be
expanded in Fourier series:
$$I(r,\phi)=I_{0}(r)+\sum_{m=1}^{\infty} I_{mc}(r)\cos m\phi + \sum_{m=1}^{\infty} I_{ms}(r)\sin m\phi$$
            or
$$I(r,\phi)=I_{0}(r)+\sum_{m=1}^{\infty} I_{m}(r)\cos
[m(\phi-\phi_{m})]~,$$ where $I_{0}(r)$ is the azimuthally-averaged
intensity in a circular annulus at a radius $r$ in the galaxy plane,
$I_{mc}$ and $I_{ms}$ are the cosine and sine amplitudes,
respectively and $\phi_{m}$ is the phase for each Fourier component
$m$.

The $I_0(r)$ ($m$=0) term defines the axisymmetric background, and
has contributions from all components, including the bulge, disk,
bar, and spiral arms. The bar and the spiral arms are
non-axisymmetric components, whose Fourier description requires a 2D
treatment in both radial and angular polar coordinates. Our 2D
analysis differs from standard 2D Fourier transforms
\citep[e.g.][]{considere88,puerari92} in that we do not transform
the whole 2D image into its frequency components, but operate on 1D
azimuthal profiles at successive radii, and use averages to derive
the radial amplitudes of different $m$ components.

The Fourier $I_{m}$ amplitudes are expressed by:
$$I_{m}=\sqrt{I_{mc}^{2}+I_{ms}^{2}} $$

\subsubsection{Bar-Spiral Separation}

The galaxies have been deprojected using the IRAF task IMLINTRAN.
The orientation parameters (mean orientation angle and axial ratio)
of the disk were input parameters in this IRAF routine. The mean
orientation angle (of the galaxy's major axis) is defined relative
to an XY plane overlapping the image of a galaxy, centered on the
galaxy's center, whose identification is explained in section 3 of
\citet{durbala08}. It is measured counterclockwise relative to the
positive x-axis. It is provided by the BUDDA
code\footnote{http://www.mpa-garching.mpg.de/~dimitri/budda.html}
\citep[BUlge/Disk Decomposition Analysis;][]{deSouza04}. We make
available the orientation measures in Table 1. The axial ratio is
tabulated as an inclination measure \emph{i}, \emph{cos(i)= b/a} in
our previous photometric study \citet{durbala08}.

We assume that the galaxy disk is circular and the deprojection is
performed about the major axis of the galaxy. For deprojection
purposes we are forced to apply a simplifying approach; in the
``face-on'' display of the galaxy the bulge should be as close as
possible to a circular shape. Thus we have three different cases: a)
in most situations, the deprojection procedure automatically leads
to a circularly shaped bulge, b) in some cases the bulge is already
circular in the original image (prior to deprojection), in which
case we subtract the bulge model given by the BUDDA decomposition
code first (see \citealt{durbala08}, we then deproject the
bulge-subtracted image and finally we add back the bulge model and
c) if neither before nor after deprojection the bulge appears
circular we outline the following recipe: 1) using BUDDA we force a
circular bulge model fit to the SDSS reduced image (before
deprojection), even though the bulge may not appear circular; 2) we
subtract the bulge model from the SDSS image; 3) we deproject the
resultant image (which is now bulge-less); 4) we add back the
adopted BUDDA bulge model from step 1 to the ``face-on'' bulgeless
deprojected image from step 3; 5) we take an average of the image
produced in step 4 and the image obtained by directly applying the
aforementioned method (a), when an elongated bulge appears after
deprojection. We emphasize that the averaging process affects only
the bulge component within the image. The resultant image in step 3
and the deprojected image (method (a)) are basically identical
outside the bulge region. In Table 1 (column 3) we indicate the
bulge deprojection method employed for each galaxy.

We are aware that the true morphology/geometry of bulges could be
far more complicated and that a round/axisymmetric bulge may be an
oversimplification. Nonetheless, our assumptions are beneficial for
two reasons: i) they do not artificially create ovals by
deprojection and ii) do not severely interfere with the study of
spiral arm morphology within galaxies. The fact that case (a) was
typical for the large majority of our galaxies (70\%) gives some
support to our simplifying assumptions about the bulge.

The first step of the Fourier analysis is bar-spiral separation. A
bar is a feature that is dominated by even Fourier terms. The bar is
separated by fitting a single or a double Gaussian function in the
bar region \citep{buta05}. In a few cases neither of the two models
appear satisfactory so the symmetry assumption is used, i.e. the
left side of the profile can be mirrored. Sometimes there is only a
single maximum intensity (e.g. Figure 3, m=6 term for KIG553
explained in the next paragraphs), so the relative Fourier
intensities decrease from the peak in a similar way as they rose to
that peak \citep{buta03}. In other cases the ``mirror-axis'' falls
in between two peaks (e.g. Figure 3, m=2 and m=4 terms). Sometimes
(but not always) a profile produced by our symmetry assumption can
closely mimic a single or a double-Gauss profile. Deciding the best
solution (Gaussian, double Gaussian or ``mirroring''/symmetry
assumption) is an iterative process driven by the visual check for
minimum residuals in the bar-subtracted image. Typically under- or
over-subtraction of the bar model would show as extra- or
deficit-light patches spatially coinciding with the outer parts of
the bar.

For all cases the first 20 terms in the Fourier expansion retain
virtually all photometric information about the galaxy, thus only
these first 20 terms are used to model the bar and galaxy light
distribution. Beyond m=20 we practically reach the background noise
level.

In simple terms the bar spatial extent could be seen as the radius
over which the bar light distribution model is non-vanishing. The
bar is defined as the sum of model fits in all even Fourier terms
over its spatial extent.

The first six figures show examples of bar fitting and bar-spiral
separation for three galaxies: KIG 550, 553 and 719, illustrating
all three possible choices for bar fitting explained (single
Gaussian, double Gaussian or symmetry assumption with no attempt to
describe analytically the profile in this latter case).

$\bullet$ \textbf{KIG 550}: Figure 1 shows the bar fitting for
galaxy KIG 550. The left panels of Figure 1 display the relative
Fourier intensities $I_{m}/I_{0}$ for the first six even Fourier
terms from m=2 to m=12 (solid line) as a function of radius. The
cross symbols show the mapping of the bar. The last term used in the
Fourier expansion to describe the bar is m=10. The bar was fitted
with a Gaussian in all even Fourier terms from m=2 to m=10. The
right panels of Figure 1 present the phase profiles $\phi_{m}$ for
the same first six even Fourier terms (m=2 to m=12).

The output images obtained from the Fourier decomposition of this
galaxy are shown in Figure 2. The upper left panel displays the
original deprojected image. ``M=0-20 SUM'' image is the sum all even
and odd Fourier terms from m=0 to m=20. This image can be regarded
as a ``Fourier-smoothed'' version of the original image
\citep{buta03}. The ``BAR + DISK'' image is the sum of the bar image
(e.g. sum of all even Fourier terms within the bar limits that have
a non-negligible contribution) and m=0 image (i.e. axisymmetric
light distribution). The ``SPIRAL + DISK'' image is the ``M=0-20
SUM'' image minus the bar image.

$\bullet$ \textbf{KIG 553}: Figure 3 presents the bar fitting for
galaxy KIG 553. The left panels show the relative Fourier intensity
amplitudes $I_{m}/I_{0}$ for the first even Fourier terms up to m=20
(solid line). The mapping of the bar is shown with cross symbols.
The right panels of Figure 3 present the phase profiles $\phi_{m}$
for the same first ten even Fourier terms (m=2 to m=20). In this
example the last term used in the Fourier expansion to describe the
bar is m=18. For the first 5 even Fourier terms (m=2 to m=10) the
symmetry assumption is used and for the next even terms (m=12 to
m=18) the bar is modeled with a Gaussian (Figure 3 - left panels).
The fact that the phase is not constant within the inner 5\arcsec\
is a deprojection effect that we could not totally eliminate.

The output Fourier images are shown in Figure 4. The designations
are the same as in Figure 2.

$\bullet$ \textbf{KIG 719}: This is one of the two galaxies in our
sample that harbors an AGN (Seyfert1 nucleus). The AGN component was
fitted by the BUDDA code and then subtracted from the original image
prior to proceed to the Fourier decomposition. The bar was fitted
with two Gaussians. The last term included in the Fourier expansion
to model the bar was m=10. The left panels of Figure 5 show the
relative Fourier intensity amplitudes $I_{m}/I_{0}$ as a function of
radius for the first six even Fourier terms from m=2 to m=12 (solid
line). The bar fitting is indicated with cross symbols. The phase
profiles $\phi_{m}$ as a function of radius for the same first six
even Fourier terms (m=2 to m=12) are displayed in the right panels
of Figure 5.

The Fourier images obtained after bar-spiral separation are
displayed in Figure 6, the designations being the same as for Figure
2.

\subsubsection{Estimation of Bar, Spiral and Total Strengths}

We employ the gravitational torque method
\citep{sanders80,combes81,buta01} to derive the bar, spiral and
total strengths for the galaxies in our sample. A constant
mass-to-light ratio is assumed. The procedure is described in detail
in \citet{buta03}. The vertical disk scaleheight is inferred from
the radial scalelength following the galaxy morphological type
dependent prescription from \citet{deGrijs98}.

The relative strength of the perturbation is calculated at each
radius $r$ in the plane of the galaxy as a force ratio:
$$Q_{T}(r)=\frac{|F_{T}(r,\phi)|_{max}}{\langle|F_{R}(r,\phi)|\rangle}~,$$ where $|F_{T}(r,\phi)|_{max}$ and $\langle|F_{R}(r,\phi)|\rangle$
are the maximum tangential force and the azimuthally averaged radial
force, respectively at a radius $r$. The strength is defined as the
maximum of the function $Q_{T}(r)$.

The bar strength ($Q_{b}$) is calculated using the ``BAR + DISK''
image, which includes all even Fourier terms contributing to the bar
plus the m=0 term, i.e. the mean axisymmetric background. The spiral
arms strength ($Q_{s}$) is determined from the ``SPIRAL + DISK''
image. The total strength of the galaxy ($Q_{g}$) is derived from
the ``M=0-20 SUM'' image (so-called ``Fourier-smoothed'' image). The
total strength includes both the bar and the spiral structure. In a
strongly barred galaxy $Q_{g} \approx Q_{b}$, while in a galaxy
where the spiral dominates $Q_{g} \approx Q_{s}$.

Figures 7, 8 and 9 present the relative strength of the
gravitational perturbation/torque $Q_{T}(r)$ as a function of radius
for KIG 550, KIG 553 and KIG 719, respectively. Bar strength
$Q_{b}$, spiral strength $Q_{s}$ and total strength $Q_{g}$ are
indicated on the figures as absolute maxima.

\section{Fourier Analysis}

Table 1 includes the Fourier-derived parameters for our sample. The
designations of each column are as follows: (1) galaxy name, (2)
orientation angle (see \S\S\S~3.1.1), (3) bulge deprojection method
(see \S\S\S~3.1.1), (4) total strength $Q_{g}$, (5) bar strength
$Q_{b}$, (6) spiral arms strength $Q_{s}$, (7) $A_{2b}$, (8)
$A_{4b}$, (9) $A_{6b}$, (10) Fourier bar length and (11) radius of
maximal bar torque r(Q$_{b}$).

We define $A_{mb}$ as the maximum of the relative Fourier intensity
amplitudes:
$$A_{mb} = \left(\frac{I_{m}}{I_{0}}\right)_{max}~,$$ where m is an even integer number.
The $A_{mb}$ indicates the contribution of the non-axisymmetric
component relative to the axi-symmetric background, thus one may see
it as a ``contrast'' measure. Hereafter we would use it as such.

For practical reasons the adopted definition for bar length
l$_{bar}$ is not fully identical to the bar spatial extent described
in \S\S\S~3.1.1. The length of the bar l$_{bar}$ is the spatial
(radial) extent where the bar model fit (section \S\S\S~3.1.1) is
non-zero AND the phase is nearly constant
\citep{laurikainen04,laurikainen05} in both m=2 and m=4 terms. Taken
independently, for the large majority of cases, the two criteria are
in agreement within 2$\sigma$ uncertainty. By ``nearly constant'' we
mean that we typically allow for a maximum of 10 degrees variation
($\pm$ 5 degrees relative to an average). This provides a rather
conservative estimate that allows us to have common grounds with the
comparison sample presented later in \S\S~4.7.

We find a very tight correlation (correlation coefficient 0.95)
between the Fourier l$_{bar}$ and the radius where the bar torque
gets the maximal value Q$_{b}$. It is shown in Figure 10 along with
the best linear regression fit. The slope of the linear fit is 1.42.
This is in good agreement with the empirical relation between
r(0.25A$_{4b}$) (i.e., the radius where the I$_{4}$/I$_{o}$(r)
profile declines to 25\% of its maximum A$_{4b}$) and r(Q$_{b}$)
proposed by \citet{buta09}. That reference reports that
r(0.25A$_{4b}$) provides a very good approximation for the visual
bar radius.

We checked whether our sample is affected in terms of Fourier
measures by biases related to inclination or redshift. We found no
correlations between the Fourier derived parameters and inclination
or redshift.

Table 2 presents mean/median measures of $Q_{g}$ and $Q_{s}$ in
three morphological bins Sb-Sbc-Sc for the whole sample of N=93
galaxies. Table 3 provides average values for the strength measures
$Q_{g}$, $Q_{s}$ and $Q_{b}$ along with the bar contrast A$_{2b}$
and bar length l$_{bar}$ for the sample of N=46 barred galaxies
split in the same three bins. Table 4 gives average $Q_{g}$ =
$Q_{s}$ values for N=47 non-barred galaxies.

\subsection{Identifying Bars with the Fourier Decomposition}

In our previous paper \citep{durbala08} 55 out of 93 galaxies in the
sample were visually classified as SAB or SB. The bulge-disk-bar
decomposition code BUDDA could fit a bar for only 48 out of the 55
SAB/SB galaxies. Within the current approach the essence of a
Fourier bar definition relies on the constancy of phase. This may
lead to some discrepancy between what Fourier decomposition defines
as the bar and what our visual evaluation (or the BUDDA
decomposition code) identifies as the bar. The most sensitive (i.e.
uncertain) cases are SAB galaxies for which an oval rather than a
clear bar is assigned visually (or with the BUDDA code), but the
phase is not constant in the Fourier terms m=2 and m=4. Part of the
discrepancy could be caused by the deprojection. The original visual
classification and the BUDDA-based decomposition are both performed
without any deprojection of images, while Fourier decomposition
requires deprojected images. Actually we find that ten SAB and one
SB galaxies do not show a constant phase in the bar region in m=2
and m=4 Fourier terms, therefore they do not have a Fourier bar
component. We would like to point out another source of uncertainty
when deciding visually or with a code like BUDDA on the
presence/absence of a bar. In galaxy KIG 652, the Fourier
decomposition reveals two widely open spiral arms in the inner
region that mimic a bar in the original image and thus could be
mistakenly classified as barred. Galaxy KIG 712 shows in its
original image an elongated ring-like structure that appears
decoupled in terms of orientation from the large disk of the galaxy.
The Fourier decomposition assimilates this structure to a
Fourier-bar associated with a constant phase. It is not clear that
the bar structure in this case is real.

The Fourier decomposition offers an additional advantage when it
comes to identifying bars in galaxies that show no clear indication
of such feature by simple visual inspection. Two galaxies that we
initially classified SA are now found to have a bar/oval in the
Fourier analysis, i.e. Fourier bars, as indicated by both the large
relative amplitude in the even terms m=2,4,6 and the constancy of
phase. All in all 46 out of 93 galaxies in our sample have a
bar/oval component separated by the Fourier analysis.

\subsection{Total Nonaxisymmetric Strength}

Figure 11 presents the distribution of the total strength for the
galaxies in our sample. Mean ($\pm$ standard deviation) and median
for the distribution are indicated on the plot. Galaxies in our
sample cover a wide range in total strengths between 0.05 and 0.55
with the bulk of the sample concentrated between 0.05 and 0.3.

Table 2 presents average values (mean and median) for the spiral and
total strength measures for all galaxies in our sample. Total
strength $Q_{g}$ decreases from Sb to Sbc and then it slightly
increases from Sbc to Sc morphological types. Tables 3 and 4 show
average strength parameters for barred and non-barred galaxies,
respectively. Barred galaxies show total strength $Q_{g}$ $\sim$ 1.5
$\times$ larger than non-barred galaxies. This trend is seen for all
morphological types in our examined range Sb-Sbc-Sc.

\subsection{Bar Strength and Bar Contrast}

Figure 12 presents the distribution of the bar torque strength for
the barred galaxies (N=46 Fourier bars) in our sample. Mean ($\pm$
standard deviation) and median for the distribution are indicated on
the figure. Barred galaxies in our sample show a wide spread in bar
strength between 0.05 and 0.55 with the majority in the range
0.15-0.25. However, on average, there is no significant difference
between the three morphological groups Sb-Sbc-Sc in terms of bar
strength (Table 3).

The average values of the maximum relative Fourier amplitudes in m =
2, 4 and 6 Fourier terms (A$_{2b}$, A$_{4b}$ and A$_{6b}$) show a
clear decline along the morphological range we focus on, with Sb
types showing the largest values and Sc types the lowest. In Table 3
we show only A$_{2b}$ average values in each morphological bin, but
not the other two bar contrast terms for m=4 and m=6 because a few
barred galaxies have a negligible bar Fourier contribution from the
4th and/or 6th term.

Figures 13(a)-(c) show the relation between bar strength Q$_{b}$ and
the maximum relative Fourier amplitudes A$_{2b}$, A$_{4b}$ and
A$_{6b}$, respectively. The three morphological types Sb-Sbc-Sc are
displayed with different symbols (see figure's legend). We see a
clear morphological separation in each panel, largely driven by
A$_{2b}$, $A_{4b}$ and $A_{6b}$. Sb galaxies tend to have larger
maximum relative Fourier amplitudes while Sc seem to show smaller
values. Sb galaxies almost always have values of Q$_{b}$ larger than
$\approx$ 0.15. Sbc-Sc galaxies seem to show a wider range in
Q$_{b}$, including values smaller than 0.15. For the plot of Q$_{b}$
versus A$_{2b}$ the best (linear) correlation coefficient is
obtained for Sbc and Sc galaxies (R=0.89 and R=0.85, respectively)
while Sb galaxies have R=0.47. We masked one point (KIG 339) when we
calculated the correlation coefficient for Sbc galaxies. Although
visual morphological classification retains some subjectivity, the
separation seen in plots like those presented in Figure 13 may be
regarded as an indirect confirmation of the robustness of
classification. Probably KIG 339 should have been classified as an
Sb instead of Sbc, since it shows as up in all panels in the space
occupied by Sb galaxies.

\subsection{Spiral Arm Strength}

Figure 14 shows the histogram distribution of the spiral strengths
Q$_{s}$ for our sample. Mean ($\pm$ standard deviation) and median
of the distribution are indicated on the plot. Galaxies in our
sample display spiral strengths between 0.05 and 0.45 with rare
cases of Q$_{s}$ $>$ 0.3.

Sc galaxies appear to show the strongest spiral structure (Table 2),
the effect being even more noticeable when restricting the
comparison to the barred subsample (Table3). In non-barred spiral
galaxies we don't see any clear trend for Q$_{s}$ (see Table 4). We
should also point out that barred and non-barred galaxies seem to
show similar spiral strengths (Table 3 versus Table 4), in contrast
to the total strength Q$_{g}$ where we noted a systematic effect,
with barred galaxies being 1.5 x stronger within each morphological
bin along the Sb-Sbc-Sc sequence (see section \S\S~4.2)

\subsection{The Interplay between Bar and Spiral Components}

Figure 15(a) shows the spiral strength as a function of bar strength
for the galaxies in our sample. The three morphological types are
indicated with different symbols (see figure's legend). No clear
correlation between spiral and bar strength is seen. Figure 15(b)
shows such a plot of spiral strength Q$_{s}$ as a function of
A$_{2b}$. Again no correlation between bar contrast and spiral
strength is revealed, but now the morphological segregation between
earlier and later types is evident. The clearest separation between
Sb and Sc types is enhanced here by the fact that Sb types show on
average the largest A$_{2b}$ and lowest Q$_{s}$ values, while Sc
galaxies show the opposite tendency (Table 3). The morphological
separation seen in panel b still holds if one tries to plot Q$_{s}$
versus A$_{4b}$ or Q$_{s}$ versus A$_{6b}$ (not shown here).

Figure 16(a) presents the total strength of the galaxy as a function
of the bar strength for the barred galaxies in our sample (N = 46).
The three morphological types are indicated with different symbols.
The solid line represents the best linear fit (correlation
coefficient R=0.96). The two parameters are very well correlated,
which is not the case for Q$_{g}$ and Q$_{s}$ in Figure 16(b). The
two panels of Figure 16 emphasize the noise in bar-spiral
separation. The strong correlation between total and bar strength
indicates that the former is a good tracer of the latter. We find
the following linear relation:
$$Q_{g}=0.829 \cdot Q_{b} + 0.079$$

It is important to note that for the barred galaxies (Table 3)
Q$_{b}$ is systematically larger than Q$_{s}$ within each
morphological segment Sb-Sbc-Sc. In most barred galaxies the total
torque is dominated by the bar contribution (Q$_{b}$ $>$ Q$_{s}$ in
34 out of 46 barred galaxies).

\subsection{The Length of Fourier Bars}

Figure 17 presents the distribution of bar lengths for the N=46
barred galaxies in our sample. Practically all barred galaxies in
our sample display bar lengths (radii) less than 10 kpc. The last
column of Table 3 gives the average values of the bar lengths for
the three morphological types represented in our sample Sb-Sbc-Sc.
The size of the bar decreases by almost a factor of $\approx$ 3 from
Sb to Sc galaxies. The decreasing trend in bar sizes is similar to
that reported in our previous paper \citep{durbala08}, with the
exception that bar sizes were found to decrease by a factor of 2
from Sb to Sc galaxies in that study. In \citet{durbala08} bar sizes
were determined from bulge-disk-bar decomposition (BUDDA code) of
the original images (without deprojecting them).

Figure 18 presents bar strength and bar contrast in the m = 2 term
(panel a and b, respectively) as a function of the Fourier bar size.
Panel a clearly indicates that the longer bars are not necessarily
the stronger ones. Panel b on the other hand shows a significant
linear correlation (correlation coefficient R=0.68) and tells us
that the longer the bar, the more prominent it appears in the sense
that it shows a bigger contrast in the m=2 Fourier term. The clear
correlation shown in panel b is preserved even when replacing the
absolute bar size l$_{bar}$ with the normalized quantity
l$_{bar}$/a$_{25}^{i}$ (linear correlation coefficient R=0.69),
where a$_{25}^{i}$ is the galactic disk semimajor axis of the 25 mag
arcsec$^{-2}$ isophote in the SDSS i-band. The morphological
separation is evident in both panels b and c with earlier Hubble
types having longer and larger relative Fourier amplitude bars in
m=2 (see also \citealt{elmegreen07}).

\subsection{Comparison with the OSU sample}

In this subsection we compare our Fourier-derived measures for our
isolated sample with similar measures for a sample selected without
isolation criteria. The best comparison sample available at this
time is the Ohio State University Bright Galaxy Survey
\citep[hereafter OSU;][]{eskridge02}. Total strengths Q$_{g}$ for
the OSU sample are available in \citet{laurikainen04} and the bar
and spiral strengths Q$_{b}$ and Q$_{s}$ are presented in
\citet{buta05}. We note that the Fourier measures for the OSU sample
are derived from H-band (near-IR) images. The OSU sample has a
comparable number of Sb-Sc galaxies (N=92 galaxies with
Fourier-derived measurements, out of 116 morphologically classified
in this narrow range). We adopted the RC3 catalogue
\citep{deVaucouleurs91} morphological classification for the OSU
sample. Both our sample and the 92 OSU galaxies show a similar
absolute magnitude M$_{B}$ distribution, ranging from -22 to -18
with a mean/median of -20.4). The number of galaxies in each
morphological type bin Sb-Sbc-Sc is also very similar to our sample
(25-32-35). We defined a subsample of N=60 barred galaxies from the
N=92 OSU Sb-Sc sample considering that a galaxy is classified as
``barred'' if it shows a Fourier bar, i.e. a constant phase in the
m=2 and m=4 terms. The Fourier bar length measurements for the OSU
sample are tabulated in \citet{laurikainen04}.

Figure 19(a) presents the histogram distribution of the total
strengths for the OSU Sb-Sc galaxies (N=92). Mean and median values
are shown on the graph. The Q$_{g}$ distributions of the OSU and our
isolated sample (recall Figure 11) are very similar, with only three
OSU galaxies exceeding Q$_{g}$$\sim$0.55. A Kolmogorov-Smirnov (KS)
test\footnote{www.nr.com} gives a 47.6\% probability of the null
hypothesis (i.e., the two samples are drawn from the same parent
population).

Figure 19(b) displays the distribution of the bar strengths Q$_{b}$
for OSU Sb-Sc barred galaxies (N=60). Again we find that the Q$_{b}$
distributions of the OSU and our isolated sample (recall Figure 12)
are very similar in terms of the range covered and average values,
with only two OSU galaxies exceeding Q$_{b}$$\sim$0.55. We note
however that the CIG/AMIGA sample of barred galaxies shows a strong
concentration (50\%) in the range Q$_{b}$=0.15-0.25, while the OSU
barred sample includes only $\sim$ 25\% in the same interval. A KS
test gives a 73.1\% probability of the null hypothesis. The
similarity between the bar strength distribution in isolated
galaxies and OSU disk galaxies is reported also in \citet{verley07b}
based on a comparison that included a broader morphological range,
i.e later than S0/a.

The spiral arm strength Q$_{s}$ distribution for the whole OSU Sb-Sc
galaxies (N=92) is presented in panel c of Figure 19. Only two OSU
galaxies show Q$_{s}$ in excess of 0.35. A KS test gives a 0.4\%
probability of the null hypothesis, which may indicate a significant
difference between the CIG/AMIGA sample (Figure 14) and OSU galaxies
in terms of spiral strength measure.

Figure 19(d) shows the histogram distribution of the Fourier bar
length l$_{bar}$ for the OSU sample of N=60 barred galaxies. This
distribution appears significantly different from that for the
CIG/AMIGA sample (Figure 17); the OSU sample is clearly lacking
large bars. A KS test confirms that the two distributions are
different, giving a 0.2\% probability of the null hypothesis.

A more detailed comparison between the CIG/AMIGA and OSU samples is
possible if one focuses on the narrow morphological types (bins)
Sb-Sbc-Sc. We present average values of the Fourier decomposition
measures for the OSU sample in Tables 5, 6 and 7 following the
framework illustrated in Tables 2, 3 and 4 for the CIG/AMIGA sample,
which facilitates a straightforward parallel analysis\footnote{We
should point out that in Table 7, Q$_{g}$ is not equal to Q$_{s}$
(as was the case for the isolated galaxies in Table 4). This is due
to a slightly different approach of the aforementioned references
that provide the Fourier parametrization for the OSU sample; the
authors include a bar component in all galaxies, thus for some
galaxies they report a nonvanishing Q$_{b}$ (typically smaller than
0.05) even though visually one cannot unambiguously identify a
bar.}.

We can summarize several differences between the isolated and the
OSU samples:

i) Comparing Q$_{s}$ for all (barred+nonbarred) we note that the
isolated Sb and Sc galaxies show larger average values relative to
OSU Sb and Sc galaxies, but the Sbc types show rather similar
Q$_{s}$ measures (Tables 2 and 5).

ii) In Table 2 (isolated galaxies) we observe a decline for the
average total strength Q$_{g}$ from Sb to Sbc, but for the OSU
sample we see a reversed trend from Sb to Sbc (Table 5; see also
Figure 14 in \citealt{buta04}).

iii) Isolated barred galaxies (Table 3) show an almost constant
Q$_{b}$ for all three morphological bins, while in the OSU sample
(Table 6) there is a slightly increasing trend from Sb through Sc.

iv) Isolated barred Sb galaxies (Table 3) show larger spiral
strength Q$_{s}$ measures than their OSU counterpart (Table 6). Sbc
and Sc barred galaxies are similar in terms of average Q$_{s}$ in
the two samples.

v) The average Q$_{g}$ for Sb isolated barred galaxies is larger
than the average Q$_{g}$ for the barred Sb from OSU (Tables 3 and
6). The OSU galaxies show an increasing trend along the Sb-Sbc-Sc
sequence, but the isolated barred galaxies show a dip at Sbc types.

vi) In terms of bar contrast measure A$_{2b}$ the isolated sample
shows a clear decline (about a factor of two) along the Sb-Sbc-Sc
morphological sequence with a larger difference between Sb and Sbc
(Table 3). The OSU sample shows very similar A$_{2b}$ averages for
Sb and Sbc bins and only a modest decline (if any) between Sbc and
Sc types (Table 6).

vii) The Fourier bar length l$_{bar}$ for the isolated sample shows
a decreasing trend from Sb through Sc, overall by about a factor
three between Sb and Sc (Table 3). However, the OSU sample shows
only slightly shorter bars for the latest types Sc, while the Sb and
Sbc are on average much more similar.

viii) Intercomparison by morphological bins reveals that the
isolated and OSU Sb barred galaxies show similar average A$_{2b}$
values (Tables 3 and 5), but for Sbc and Sc types OSU galaxies show
larger values. For Sb and Sbc types, the bars in isolated galaxies
are systematically longer, but in the case of Sc types there is no
significant difference. As shown in Figure 20 in both samples
CIG/AMIGA and OSU there is a positive trend between the bar contrast
and its size. The isolated barred galaxies apparently show a
different scaling relation between l$_{bar}$ and A$_{2b}$ than the
barred galaxies from the OSU sample within the same morphological
interval T=3-5. For a similar l$_{bar}$ the isolated galaxies show a
lower contrast. However this difference can be attributed to the
fact that we perform our analysis on SDSS i-band images and OSU
Fourier measures are extracted from H-band near-IR images. It is
well known that near-IR images, much less affected by extinction and
good tracers of old stellar populations, could reveal more clearly
the presence/absence of bars. This is also reflected by the
significantly larger number of barred galaxies in the comparison OSU
sample (60 out of 92).

\subsection{Spiral Arm Multiplicity}

Using the $I_{mc}$ and $I_{ms}$ amplitudes we could reconstruct the
images of the individual \emph{m} Fourier terms. For example, the
m=1 image would be given by $I_{1c}(r)\cos \phi + I_{1s}(r)\sin
\phi$ and the m=2 image would be given by $I_{2c}(r)\cos 2\phi +
I_{2s}(r)\sin 2\phi$, etc. (see the first equation in \S\S~3.1).

Figure 21 displays a concrete example; it shows the reduced and
deprojected SDSS i-band image of KIG 281 and the reconstructed  m =
1, 2, 3, 4, 5 Fourier term images. KIG 281 has two symmetric spiral
arms ($\cos 2\phi$ periodicity), therefore the dominant Fourier term
is m=2. From a practical point of view, the Fourier terms with a
nontrivial contribution to the spiral structure of a galaxy are
those that match visually observable features in the deprojected
image. In all cases the spiral structure is fully reconstructed
without including terms beyond m=6 and in most cases the first three
terms suffice.

In this subsection we consider for analysis only 86 galaxies. Seven
out of 93 galaxies do not show clear spiral arm morphology in their
images. Therefore, we exclude them from the analysis of the m = 1-6
Fourier term images performed in this subsection. Table 8 offers a
census of spiral arm multiplicity encountered among the N=86 sample
of isolated galaxies that are subject to Fourier analysis.

About 40\% of the galaxies in our sample (N=86) have \textbf{only} a
two-armed pattern (m=2), $\sim$ 4\% have \textbf{only} a three-armed
pattern (m=3) and $\sim$ 1\% have \textbf{single} m=1 spiral arms.

About 87\% of our galaxies harbor m=2 spiral arms, $\sim$ 38\% have
m=3 spiral arms and $\sim$ 20\% host m=1 spiral arms. 13\% of the
galaxies have both m=1 and m=2 spiral arms. About 28\% of the
galaxies in our sample have both m=2 and m=3 spiral arms, with the
two-armed pattern usually in the inner part of the galaxy and m=3
spiral arms in the outer part. A representative example in this
sense would be KIG 260. Figure 22 displays the SDSS i-band image of
KIG 260 and the m=1, 2, 3 Fourier terms images. One could easily
notice two inner spiral arms (m=2) starting at the end of the bar
and three spiral arms (m=3) in the outer part of the galaxy.

A particularly intriguing case is galaxy KIG 652. It was classified
as SAB in our previous paper \citep{durbala08}. The best
bulge-disk-bar decomposition solution returned by the BUDDA included
a bar component for this galaxy. The reconstruction of the m=1-6
Fourier terms revealed that the bar is not real and in fact there
are two counter-winding inner spiral arms that mimic a bar-like
feature in an image that is not deprojected (as used by the BUDDA
code). Figure 23 displays the reduced and deprojected SDSS i-band
image and the m = 1, 2, 3 Fourier images. In the m=2 image one can
see the two inner counter-winding spiral arms (very open). The m=3
image shows the three outer spiral arms. KIG 652 has m=2 and m=3
spiral arms winding in opposite directions. The m = 3 Fourier images
of KIG 260 and KIG 652 (Figures 22 and 23, respectively) show
possible counterwinding spiral structure in their inner regions.
However, within resolution and deepness constraints we cannot
confirm those structures by direct visual inspection of the
deprojected SDSS i-band images.

Another interesting case is KIG 282, whose deprojected SDSS image is
shown in Figure 24 along with the Fourier reconstructed images
corresponding to m=1 through 3 terms. KIG 282 is a barred galaxy
that displays both m=2 and m=3 spiral arm morphology. It is rather
rare to see that a spiral arm in the m=3 image originates very close
to the bulge making a $\sim$45$^{o}$ angle with the bar. The other
two arms of the m=3 term show a smooth continuity with the m=2 arms,
which appear joined to the end regions of the bar.

\section{Discussion}

We have reported here the results of a Fourier decomposition
analysis for a representative sample of Sb-Sc isolated (CIG/AMIGA)
galaxies. This complements our earlier surface photometric analysis
\citep{durbala08} for the same sample. Our primary goal has been to
characterize the structural properties of galaxies likely to have
been least affected by external stimuli. The most common (2/3) kind
of isolated galaxy appears to be the late-type spiral (Sb-Sc). This
minimal-nurture sample can provide important clues about the
formation, evolution and interplay of galactic components without
the confusion added by external influences. We have focused here on
measures involving the bar and spiral arm components. We now
consider the main results of this paper in the light of some
theoretical predictions and by comparing them to other samples of
disk galaxies selected without isolation criteria.

\subsection{Properties of Bars}

Our Fourier analysis reveals that about 50\% of our sample are
barred spirals. We tested whether the barred and non-barred
subsamples are different in terms of isolation (isolation
parameters, i.e., tidal strengths for AMIGA galaxies were quantified
in \citealt{verley07a}), absolute magnitude M$_{i}$, size
a$^{i}_{25}$ and color (g-i) (tabulated in \citealt{durbala08}). We
find no statistical difference between barred and non-barred
galaxies in terms of isolation measures. This is in agreement with
the recent study of \citet{li09}, where they report no clustering
differences between barred and non-barred galaxies based on a large
sample of n=675 SDSS spiral galaxies. The barred and nonbarred
galaxies in our sample are very similar in absolute magnitude and
size, the only statistically significant difference is found for the
color (g-i), median colors are 0.88 and 0.72 for barred and
unbarred, respectively. This is probably expected given the observed
tendency of stellar bars to show higher contrast in red and near-IR
filters (see section 3.1 of \citealt{kormendy04} and references
therein). And this is also tied to the fact that the color gets
bluer from Sb to Sc, the earlier bin (Sb) having the largest
fraction of barred galaxies \citep{durbala08}.

Various studies attribute the term ``strength'' for different bar
measures, e.g. \citet{athanassoula03} refers as ``strength'' to a
measure S$_{B}$ more similar to our ``contrast'' terms A$_{mb}$,
defined in \S~4. Simulation studies
\citep[e.g.][]{athanassoula02,athanassoula03} predict an
anticorrelation between S$_{B}$ and the bar pattern speed.
\citet{sellwood00} suggested that within a disk galaxy the spiral
component can transfer material to the bar, thus making it longer
and reducing its pattern speed. This is also suggested by more
recent simulations \citep[e.g.][]{martinez06}. From these two major
theoretical conclusions it could be inferred that the longer a bar
becomes, the larger A$_{mb}$ gets. Our results do confirm such
theoretical predictions. We find that although the longer bars are
not necessarily stronger (in terms of our Q$_{b}$ torque) than the
shorter ones (Figure 18a), the longer bars show higher contrast,
i.e. there is a positive correlation between A$_{2b}$ (maximum
Fourier relative amplitudes in m=2) and Fourier bar length l$_{bar}$
(Figures 18b,c). This is also seen in the OSU sample (Figure 20).
The fact that our observed correlations in Figures 18b,c are not
very strong (correlation coefficients $\sim$0.7) is in very good
agreement with the numerical simulations that show a wide possible
range of exchanged angular momentum between galactic components
\citep{athanassoula03}.

The role of gas in the process of bar formation, growth and
interaction with the other major galactic components is still being
debated \citep[e.g.][]{berentzen07}. By transferring angular
momentum a bar can contribute to the build-up of central mass
concentrations, which in turn could lead to a declining bar
\citep{pfenniger90,norman96}, but probably not to the extent of
complete destruction \citep{shen04,athanassoula05,bournaud05}. The
interpretation of observational results is further complicated by
considering the role of the so-called ``buckling instability''
\citep[e.g.][]{debattista06,martinez06}, which could weaken the bar
within 2-3 Gyr of its formation.

Moreover, bars in gas rich spiral galaxies might be short lived
structures and in typical Sb-Sc galaxies a bar can practically
dissolve in 2 Gyr \citep{bournaud02}. This is smaller than the time
scale over which our isolated galaxies have not been visited by a
similar size neighbor, $\sim$ 3 Gyr \citep{verdes05}. The presence
of gas in galactic disks is responsible for both the destruction and
renewal of bars when the gas is accreted from outside the disk
\citep{bournaud02,block02}. Simulations with sufficient resolution
allow one to see the cyclic process of formation, destruction and
reformation of bars \citep{heller07}.

According to \citet{block02} the fate of pure isolated disks (i.e.
closed systems that do not accrete mass from outside) is that they
``would become unbarred and their spiral structure would disappear;
many disks would then be nearly axisymmetric after a few Gyr''.
\citet{block02} argue that the observed strength (torque)
distribution for disk galaxies with a striking depression at low
values and an extended tail at large values\footnote{Note that
\citet{block02} employ the OSU sample using a constant radial to
vertical h$_{r}$/h$_{z}$ = 12 ratio for all morphological types.}
can be accounted for only by considering that spiral galaxies are
open systems, actively and \emph{continuously} accreting mass today
(see also \citealt{sellwood84}). The origin of the accreted gas is
not considered, but it appears that accretion of dwarf satellites is
far from enough in their simulations. The CIG/AMIGA isolated
galaxies also lack large companions by definition. In the light of
such arguments, one can conclude that the accreted matter must come
from either some sort of galactic internal reservoirs or from
intergalactic cosmic filaments \citep{combes08}. A very recent study
\citep{bekki09} investigates, using numerical simulations, ``whether
and how stellar winds from bulges (or stellar ejecta due to
supernova feedback) can be accreted onto the disks after
hydrodynamical interaction with the gaseous halos''. Although that
study explores a chemical connection between bulge and disk
components, it certainly proposes a viable mechanism to add new mass
onto the disks.

The fate of bars can be significantly affected by tidal interactions
\citep[e.g.][]{noguchi87,gerin90,miwa98,berentzen03,berentzen04}. We
consider that CIG/AMIGA are minimally affected by external
interactions. We looked for trends/correlations between tidal
strengths \citep{verley07a} and estimated bar, spiral and total
torque strength parameters for the galaxies in our sample and found
none. We should also point out that no correlation was found between
the basic structural parameters of the bulge, disk and bar presented
in \citet{durbala08} and the tidal strength measures quantified in
\citet{verley07a}.

We find that Q$_{b}$ and l$_{bar}$ do not correlate for our
CIG/AMIGA sample. We also explore this torque-bar length relation by
comparing our isolated galaxies with the OSU sample (Tables 3 and
6). Even though the CIG/AMIGA galaxies host longer bars\footnote{The
isolated galaxies show larger bars than OSU galaxies both in terms
of Fourier bars analyzed herein on deprojected images and also in
terms of bar size derived from 2D light decomposition of projected
images \citep{durbala08}.} (the difference being most noticeable for
Sb and Sbc types) we do not find stronger Q$_{b}$ measures for the
CIG/AMIGA isolated galaxies.

The observed low occurrence of strong bars in both CIG/AMIGA and OSU
(see Figures 12 and 19b) may indicate either that strong bars are
very transient and/or they are allowed only by special conditions
\citep{buta05}, apparently not sampled by either of the two samples.

\subsection{Bar-Spiral Connection}

We find that in $\sim$ 74\% of the barred galaxies the strength of
bars dominates over the spiral arm strength (Table 1). This is also
seen in Table 3 where within each morphological bin Q$_{b}$ $>$
Q$_{s}$ in isolated galaxies and in Table 6 for the OSU sample. We
find that in our sample Q$_{g}$ is a very reliable tracer of the bar
strength Q$_{b}$ (Figure 16a).

A very recent study \citep{buta09} has examined on empirical grounds
the connection between the torque strength of bars and spiral
structure using near-IR K$_{s}$-band images for 23 galaxies that are
morphologically diverse. They find weak but definite indications
that stronger spirals are associated with stronger bars (see also
\citealt{block04}); their correlation is relevant for Q$_{b}$ $>$
0.3. Perhaps the energy and angular momentum exchange due to
resonance coupling between bar and spiral components
\citep{tagger87,sygnet88} is reflected in a Q$_{b}$ - Q$_{s}$
correlation only for this restricted Q$_{b}$ $>$ 0.3 regime.

Our data do not show any trend or correlation between the two
measures Q$_{b}$ and Q$_{s}$ (Figure 15(a)). However, our sample
includes only 13 (out of 46 barred) galaxies with strong Q$_{b}$ $>$
0.3 measures. From this point of view, in the isolated galaxies
investigated here bars and spirals appear to be more independent
features (see also \citealt{sellwood88}).

\subsection{Properties of Spiral Arms}

It is worth noting that in Figure 15(b), where we plot spiral
strength Q$_{s}$ versus bar contrast A$_{2b}$, we see a clear
morphological separation, although no correlation is observed in
this plot either. We find that bar strength and bar contrast (Figure
13 (a)-(c)) are very well correlated in Sbc-Sc types (see section
\S\S~4.3), but the Sb galaxies depart from that correlation along
the abscissa and they spread over a larger bar contrast range. It is
also worth indicating that on average the Sb galaxies show the
largest differences in almost all Fourier measures when comparing
isolated and OSU galaxies.

Fourier decomposition can reveal surprising cases of counter-winding
spiral structure (KIG 652 / NGC 5768). Only a few other similar
cases are known in literature: NGC 4622 \citep{bbf03}, ESO 297-27
\citep{grouchy08}, NGC 3124 \citep{purcell98,buta99} and IRAS
18293-3413 \citep{vaisanen08}.

We would like at this point to evaluate the relative frequency of
certain spiral arm multiplicities in our sample of isolated Sb-Sc
galaxies in contrast to other similar studies. The only reference
where a study of spiral arm multiplicity is available is the Catalog
of Southern Ringed Galaxies (CSRG; \citealt{buta95}). However, one
should keep in mind that the CSRG galaxies were evaluated in terms
of such multiplicities by direct visual inspection of their images,
without any Fourier analysis or prior deprojection of images. CSRG
is a special catalog in itself being a collection of ``ringed''
galaxies. This is why we caution the reader that any inference we
make in the light of the comparison of our sample against CSRG could
be seen as speculative for the time being. Using the on-line access
to CSRG through VizieR\footnote{http://vizier.cfa.harvard.edu} we
extracted from CSRG only the Sb-Sc galaxies, i.e. morphological
types T=3-5. We considered both the full sample thus obtained, but
also a more ``restricted'' subset imposing the conditions explained
in \citet{buta95} (relative to his Table 8). This latter subset is
also considered more reliable for statistical purposes.

Two-armed spiral patterns are the most frequent among isolated Sb-Sc
galaxies ($\sim$ 40\%). Among the Sb-Sc of the CSRG the fraction of
m = 2 is 31-33 \% and still the most frequent mode. However, large
differences are noted for m = 2 \& 3 spiral arm multiplicity. We
find in our sample 24 out of 86 m = 2 \& 3 galaxies (28\%). The
CSRG-based comparison sample includes 6-8\% such cases. However, we
note that the definitions employed by \citet{buta95} are not the
same ones applied herein (i.e. what we call here 2 \& 3 would most
likely be equivalent to 1+2, 2+1 and 3 altogether in that
reference). We cannot assess at this time whether the rarity of 2 \&
3 multiplicity combination is due to the special nature of that CSRG
catalog or it is a phenomenon more likely to occur in isolated
galaxies. But it is particularly interesting to indicate here that
the high rate of occurrence of m = 2 \& 3 combination among
CIG/AMIGA galaxies may be linked to their isolation \citep{eem92}.
The formation of strong three-arm structures may require long
episodes without strong tidal perturbations (\emph{``Perhaps
three-arm structures will provide a good measure of the time that
has elapsed since a tidal interaction''} - \citealt{eem92}).
Moreover, the fact that the Q$_{s}$ distribution for CIG/AMIGA is
significantly different than that of OSU (\S\S~4.7) may be tied to
the isolation, too.

\section{Conclusions}

Our Fourier decomposition analysis applied to a representative
sample of n$\sim$100 isolated CIG/AMIGA galaxies allows several
important conclusions:\\
$\bullet$ both the length (l$_{bar}$) and the contrast (e.g.
A$_{2b}$) of the Fourier bars decrease along the morphological
sequence Sb-Sbc-Sc, with bars in earlier types being longer and
showing higher contrast;\\
$\bullet$ a tight correlation between the bar strength Q$_{b}$ and
the bar contrast (e.g. A$_{2b}$; Figure 13a) is evident for Sbc-Sc
types, while Sb galaxies seem to depart from the trend, being
clearly separated in bar contrast measures;\\
$\bullet$ longer bars are not necessarily stronger (as indicated by
the torque measures), but longer bars show higher Fourier contrast (i.e. relative amplitudes),
in very good agreement with theoretical predictions;\\
$\bullet$ bar and spiral galactic components are independent in the
sense that the dynamical torque-strengths of the two
components are not correlated;\\
$\bullet$ the total strength Q$_{g}$ is a very reliable tracer of
the bar strength Q$_{b}$;\\
$\bullet$ for the large majority of the barred galaxies in our
sample ($\sim$ 74\%) the strength of the bar dominates over the
spiral arm strength (Q$_{b}$ $>$ Q$_{s}$), which is also noted in
the OSU comparison sample;\\
$\bullet$ barred and non-barred galaxies show similar spiral arm
strengths Q$_{s}$, while the total non-axisymmetric strength Q$_{g}$
is about 1.5 $\times$ larger in barred relative to the non-barred
galaxies (in each morphological bin Sb-Sbc-Sc);\\
$\bullet$ comparison with samples of galaxies of the same
morphological types defined and selected without isolation criteria
(e.g. OSU sample) indicates that the isolated CIG/AMIGA galaxies
host longer Fourier bars and possibly have a different distribution
of spiral torque strength Q$_{s}$;\\
$\bullet$ Fourier decomposition can reveal surprisingly rare cases
of counterwinding spiral structure (e.g. KIG 652/NGC 5768);\\
$\bullet$ our sample of isolated Sb-Sc galaxies is dominated by m=2
spiral arm multiplicity ($\sim$ 40\%);\\
$\bullet$ m = 2 \& 3 spiral arm components appear present in $\sim$
28\% of our sample and this rather large rate of occurrence may
indicate a long time without external tidal perturbations
\citep{eem92}.

\section*{Acknowledgments}

AD and RB acknowledge support of NSF Grant AST 05-07140.

This study has made use of SDSS Data Release 6. Funding for the SDSS
and SDSS-II has been provided by the Alfred P. Sloan Foundation, the
Participating Institutions, the National Science Foundation, the
U.S. Department of Energy, the National Aeronautics and Space
Administration, the Japanese Monbukagakusho, the Max Planck Society,
and the Higher Education Funding Council for England. The SDSS Web
Site is http://www.sdss.org/. The SDSS is managed by the
Astrophysical Research Consortium for the Participating
Institutions. The Participating Institutions are the American Museum
of Natural History, Astrophysical Institute Potsdam, University of
Basel, University of Cambridge, Case Western Reserve University,
University of Chicago, Drexel University, Fermilab, the Institute
for Advanced Study, the Japan Participation Group, Johns Hopkins
University, the Joint Institute for Nuclear Astrophysics, the Kavli
Institute for Particle Astrophysics and Cosmology, the Korean
Scientist Group, the Chinese Academy of Sciences (LAMOST), Los
Alamos National Laboratory, the Max-Planck-Institute for Astronomy
(MPIA), the Max-Planck-Institute for Astrophysics (MPA), New Mexico
State University, Ohio State University, University of Pittsburgh,
University of Portsmouth, Princeton University, the United States
Naval Observatory, and the University of Washington.

This research has made use of the VizieR catalogue access tool, CDS,
Strasbourg, France.

The authors kindly thank the reviewer for many helpful comments and
suggestions.

\clearpage
\begin{table}
\setcounter{table}{0}
\begin{minipage}{12cm}
\caption{Fourier-derived Parameters in i-band for the CIG/KIG
Galaxies in our Sample}
\end{minipage}
\begin{tabular}{lccccccccrr}

\hline\hline

(1) & (2) & (3) & (4) & (5) & (6) & (7) & (8) & (9) & (10) & (11) \\
Galaxy & orientation &  bulge & Q$_{g}$ & Q$_{b}$ & Q$_{s}$ & A$_{2b}$ & A$_{4b}$ & A$_{6b}$ & l$_{bar}$ & r(Q$_{b}$)\\
Name   &  angle ($^\circ$)  & method &    &     &         &          &          &         & (arcsec) & (arcsec)\\

\hline
KIG 011 &   77  &   a   &   0.091   $\pm$   0.040   &               &   0.091   $\pm$   0.040   &       &       &       &       &       \\
KIG 033 &   178 &   c   &   0.183   $\pm$   0.082   &               &   0.183   $\pm$   0.082   &       &       &       &       &       \\
KIG 056 &   71  &   a   &   0.293   $\pm$   0.023   &   0.227   $\pm$   0.023   &   0.166   $\pm$   0.022   &   0.529   &   0.255   &   0.151   &   19.0    &   11.0    \\
KIG 187 &   99  &   a   &   0.134   $\pm$   0.008   &   0.080   $\pm$   0.002   &   0.126   $\pm$   0.080   &   0.175   &   0.053   &       &   8.0 &   5.5 \\
KIG 198 &   72  &   a   &   0.176   $\pm$   0.020   &   0.108   $\pm$   0.001   &   0.171   $\pm$   0.015   &   0.217   &       &       &   12.0    &   7.0 \\
KIG 203 &   173 &   b   &   0.136   $\pm$   0.052   &               &   0.136   $\pm$   0.052   &       &       &       &       &       \\
KIG 217 &   172 &   a   &   0.183   $\pm$   0.010   &               &   0.183   $\pm$   0.010   &       &       &       &       &       \\
KIG 222 &   46  &   a   &   0.236   $\pm$   0.055   &   0.184   $\pm$   0.055   &   0.165   $\pm$   0.003   &   0.282   &   0.092   &   0.043   &   11.0    &   8.0 \\
KIG 232 &   46  &   a   &   0.391   $\pm$   0.094   &               &   0.391   $\pm$   0.094   &       &       &       &       &       \\
KIG 238 &   178 &   a   &   0.358   $\pm$   0.074   &   0.286   $\pm$   0.004   &   0.235   $\pm$   0.089   &   0.697   &   0.405   &   0.202   &   12.0    &   8.0 \\
KIG 241 &   178 &   a   &   0.260   $\pm$   0.080   &               &   0.260   $\pm$   0.080   &       &       &       &       &       \\
KIG 242 &   102 &   a   &   0.150   $\pm$   0.032   &               &   0.150   $\pm$   0.032   &       &       &       &       &       \\
KIG 258 &   34  &   b   &   0.214   $\pm$   0.042   &   0.205   $\pm$   0.024   &   0.115   $\pm$   0.038   &   0.439   &   0.125   &   0.071   &   9.0 &   8.0 \\
KIG 260 &   124 &   a   &   0.190   $\pm$   0.038   &   0.156   $\pm$   0.011   &   0.178   $\pm$   0.062   &   0.155   &   0.034   &       &   8.0 &   4.0 \\
KIG 271 &   159 &   c   &   0.156   $\pm$   0.047   &               &   0.156   $\pm$   0.047   &       &       &       &       &       \\
KIG 281 &   130 &   a   &   0.095   $\pm$   0.011   &               &   0.095   $\pm$   0.011   &       &       &       &       &       \\
KIG 282 &   135 &   a   &   0.234   $\pm$   0.022   &   0.230   $\pm$   0.015   &   0.175   $\pm$   0.050   &   0.277   &   0.053   &   0.024   &   8.5 &   4.0 \\
KIG 287 &   61  &   c   &   0.230   $\pm$   0.027   &   0.220   $\pm$   0.019   &   0.114   $\pm$   0.021   &   0.364   &   0.117   &   0.058   &   15.0    &   7.0 \\
KIG 292 &   39  &   a   &   0.307   $\pm$   0.065   &               &   0.307   $\pm$   0.065   &       &       &       &       &       \\
KIG 298 &   110 &   a   &   0.202   $\pm$   0.031   &   0.167   $\pm$   0.013   &   0.127   $\pm$   0.016   &   0.417   &   0.200   &   0.103   &   14.0    &   10.5    \\
KIG 302 &   5   &   a   &   0.443   $\pm$   0.078   &               &   0.443   $\pm$   0.078   &       &       &       &       &       \\
KIG 314 &   110 &   a   &   0.137   $\pm$   0.045   &               &   0.137   $\pm$   0.045   &       &       &       &       &       \\
KIG 325 &   53  &   a   &   0.151   $\pm$   0.061   &               &   0.151   $\pm$   0.061   &       &       &       &       &       \\
KIG 328 &   14  &   c   &   0.170   $\pm$   0.021   &               &   0.170   $\pm$   0.021   &       &       &       &       &       \\
KIG 330 &   173 &   a   &   0.172   $\pm$   0.089   &               &   0.172   $\pm$   0.089   &       &       &       &       &       \\
KIG 336 &   54  &   a   &   0.332   $\pm$   0.043   &   0.330   $\pm$   0.026   &   0.059   $\pm$   0.005   &   0.564   &   0.333   &   0.230   &   25.0    &   16.5    \\
KIG 339 &   167 &   a   &   0.180   $\pm$   0.008   &   0.175   $\pm$   0.012   &   0.092   $\pm$   0.045   &   0.833   &   0.357   &   0.193   &   19.0    &   12.0    \\
KIG 351 &   121 &   a   &   0.509   $\pm$   0.043   &   0.504   $\pm$   0.018   &   0.079   $\pm$   0.038   &   0.616   &   0.289   &   0.166   &   11.0    &   8.0 \\
KIG 365 &   99  &   a   &   0.314   $\pm$   0.033   &   0.269   $\pm$   0.018   &   0.203   $\pm$   0.069   &   0.292   &   0.106   &   0.050   &   8.5 &   5.0 \\
KIG 366 &   113 &   a   &   0.338   $\pm$   0.052   &   0.311   $\pm$   0.051   &   0.168   $\pm$   0.063   &   0.560   &   0.220   &   0.100   &   19.0    &   11.0    \\
KIG 367 &   81  &   a   &   0.139   $\pm$   0.053   &               &   0.139   $\pm$   0.053   &       &       &       &       &       \\
KIG 368 &   47  &   a   &   0.220   $\pm$   0.053   &               &   0.220   $\pm$   0.053   &       &       &       &       &       \\
KIG 386 &   136 &   a   &   0.181   $\pm$   0.050   &   0.137   $\pm$   0.027   &   0.164   $\pm$   0.057   &   0.209   &   0.069   &   0.028   &   5.5 &   4.0 \\
KIG 397 &   148 &   b   &   0.207   $\pm$   0.077   &               &   0.207   $\pm$   0.077   &       &       &       &       &       \\
KIG 399 &   14  &   a   &   0.199   $\pm$   0.024   &               &   0.199   $\pm$   0.024   &       &       &       &       &       \\
KIG 401 &   132 &   c   &   0.286   $\pm$   0.019   &               &   0.286   $\pm$   0.019   &       &       &       &       &       \\
KIG 405 &   101 &   a   &   0.199   $\pm$   0.084   &   0.094   $\pm$   0.015   &   0.200   $\pm$   0.084   &   0.134   &       &       &   3.0 &   1.0 \\
KIG 406 &   32  &   a   &   0.156   $\pm$   0.055   &               &   0.156   $\pm$   0.055   &       &       &       &       &       \\
KIG 409 &   122 &   a   &   0.291   $\pm$   0.061   &   0.210   $\pm$   0.014   &   0.293   $\pm$   0.056   &   0.201   &   0.045   &       &   4.0 &   2.0 \\
KIG 410 &   5   &   a   &   0.238   $\pm$   0.128   &               &   0.238   $\pm$   0.128   &       &       &       &       &       \\
KIG 429 &   124 &   a   &   0.183   $\pm$   0.064   &               &   0.183   $\pm$   0.064   &       &       &       &       &       \\
KIG 444 &   25  &   a   &   0.268   $\pm$   0.078   &   0.180   $\pm$   0.002   &   0.266   $\pm$   0.078   &   0.171   &       &       &   2.5 &   2.0 \\
KIG 446 &   154 &   b   &   0.091   $\pm$   0.031   &               &   0.091   $\pm$   0.031   &       &       &       &       &       \\
KIG 460 &   168 &   c   &   0.170   $\pm$   0.027   &   0.169   $\pm$   0.010   &   0.122   $\pm$   0.010   &   0.142   &   0.061   &   0.044   &   5.5 &   3.0 \\
KIG 466 &   52  &   a   &   0.396   $\pm$   0.054   &   0.388   $\pm$   0.011   &   0.134   $\pm$   0.042   &   0.277   &   0.100   &   0.042   &   11.8    &   1.0 \\
KIG 489 &   178 &   a   &   0.249   $\pm$   0.100   &               &   0.249   $\pm$   0.100   &       &       &       &       &       \\
KIG 491 &   39  &   a   &   0.074   $\pm$   0.010   &               &   0.074   $\pm$   0.010   &       &       &       &       &       \\
KIG 494 &   77  &   a   &   0.277   $\pm$   0.136   &   0.241   $\pm$   0.017   &   0.241   $\pm$   0.082   &   0.220   &   0.032   &       &   4.0 &   1.0 \\
KIG 499 &   72  &   b   &   0.261   $\pm$   0.033   &   0.239   $\pm$   0.011   &   0.140   $\pm$   0.043   &   0.539   &   0.217   &   0.090   &   10.2    &   7.0 \\
KIG 502 &   15  &   b   &   0.119   $\pm$   0.019   &               &   0.119   $\pm$   0.019   &       &       &       &       &       \\
KIG 508 &   80  &   a   &   0.404   $\pm$   0.088   &   0.370   $\pm$   0.005   &   0.254   $\pm$   0.119   &   0.555   &   0.222   &   0.051   &   6.0 &   2.0 \\
KIG 512 &   136 &   a   &   0.382   $\pm$   0.068   &   0.368   $\pm$   0.016   &   0.187   $\pm$   0.037   &   0.364   &   0.140   &   0.052   &   17.0    &   3.0 \\
KIG 515 &   140 &   a   &   0.161   $\pm$   0.037   &   0.108   $\pm$   0.004   &   0.161   $\pm$   0.037   &   0.076   &       &       &   4.5 &   1.0 \\
KIG 520 &   85  &   a   &   0.075   $\pm$   0.014   &   0.058   $\pm$   0.020   &   0.075   $\pm$   0.014   &   0.100   &       &       &   3.5 &   2.0 \\
KIG 522 &   107 &   a   &   0.306   $\pm$   0.021   &   0.205   $\pm$   0.019   &   0.188   $\pm$   0.038   &   0.357   &   0.116   &   0.038   &   5.0 &   3.0 \\
KIG 525 &   34  &   a   &   0.231   $\pm$   0.022   &   0.199   $\pm$   0.001   &   0.189   $\pm$   0.037   &   0.389   &   0.233   &   0.128   &   12.2    &   9.0 \\
KIG 532 &   16  &   a   &   0.492   $\pm$   0.148   &   0.490   $\pm$   0.013   &   0.222   $\pm$   0.081   &   0.528   &   0.185   &   0.074   &   6.0 &   2.0 \\
KIG 550 &   57  &   a   &   0.244   $\pm$   0.025   &   0.223   $\pm$   0.001   &   0.143   $\pm$   0.037   &   0.432   &   0.214   &   0.123   &   15.0    &   10.0    \\
KIG 553 &   50  &   a   &   0.192   $\pm$   0.010   &   0.132   $\pm$   0.011   &   0.140   $\pm$   0.009   &   0.562   &   0.300   &   0.206   &   20.0    &   11.0    \\
KIG 560 &   161 &   a   &   0.252   $\pm$   0.024   &   0.230   $\pm$   0.012   &   0.184   $\pm$   0.058   &   0.238   &   0.044   &       &   4.5 &   1.0 \\
KIG 571 &   125 &   b   &   0.103   $\pm$   0.028   &               &   0.103   $\pm$   0.028   &       &       &       &       &       \\
KIG 575 &   52  &   a   &   0.080   $\pm$   0.036   &               &   0.080   $\pm$   0.036   &       &       &       &       &       \\

\end{tabular}

\end{table}

\clearpage
\begin{table}
\textbf{Table 1.}--continued

\centering
\begin{tabular}{lccccccccrr}

\hline\hline
(1) & (2) & (3) & (4) & (5) & (6) & (7) & (8) & (9) & (10) & (11) \\
Galaxy & orientation &  bulge & Q$_{g}$ & Q$_{b}$ & Q$_{s}$ & A$_{2b}$ & A$_{4b}$ & A$_{6b}$ & l$_{bar}$ & r(Q$_{b}$)\\
Name   &  angle ($^\circ$)  & method &    &     &         &          &          &         & (arcsec) & (arcsec)\\

\hline
KIG 580 &   132 &   a   &   0.149   $\pm$   0.054   &               &   0.149   $\pm$   0.054   &       &       &       &       &       \\
KIG 598 &   62  &   c   &   0.250   $\pm$   0.043   &               &   0.250   $\pm$   0.043   &       &       &       &       &       \\
KIG 612 &   103 &   a   &   0.205   $\pm$   0.014   &   0.189   $\pm$   0.001   &   0.091   $\pm$   0.018   &   0.473   &   0.244   &   0.108   &   10.8    &   7.0 \\
KIG 626 &   155 &   a   &   0.292   $\pm$   0.142   &   0.279   $\pm$   0.011   &   0.288   $\pm$   0.060   &   0.225   &   0.057   &       &   10.0    &   3.0 \\
KIG 630 &   64  &   a   &   0.175   $\pm$   0.062   &               &   0.175   $\pm$   0.062   &       &       &       &       &       \\
KIG 633 &   70  &   a   &   0.113   $\pm$   0.069   &               &   0.113   $\pm$   0.069   &       &       &       &       &       \\
KIG 639 &   20  &   b   &   0.139   $\pm$   0.034   &               &   0.139   $\pm$   0.034   &       &       &       &       &       \\
KIG 640 &   30  &   a   &   0.084   $\pm$   0.036   &               &   0.084   $\pm$   0.036   &       &       &       &       &       \\
KIG 641 &   45  &   a   &   0.225   $\pm$   0.018   &   0.197   $\pm$   0.003   &   0.113   $\pm$   0.034   &   0.409   &   0.178   &   0.107   &   12.0    &   9.0 \\
KIG 645 &   0   &   b   &   0.145   $\pm$   0.036   &               &   0.145   $\pm$   0.036   &       &       &       &       &       \\
KIG 652 &   120 &   a   &   0.217   $\pm$   0.068   &               &   0.217   $\pm$   0.068   &       &       &       &       &       \\
KIG 665 &   70  &   a   &   0.093   $\pm$   0.048   &               &   0.093   $\pm$   0.048   &       &       &       &       &       \\
KIG 671 &   117 &   c   &   0.362   $\pm$   0.052   &   0.336   $\pm$   0.003   &   0.161   $\pm$   0.100   &   1.003   &   0.561   &   0.349   &   16.5    &   10.5    \\
KIG 689 &   80  &   a   &   0.280   $\pm$   0.144   &   0.240   $\pm$   0.006   &   0.137   $\pm$   0.095   &   0.200   &   0.045   &       &   6.0 &   1.0 \\
KIG 712 &   152 &   b   &   0.412   $\pm$   0.093   &   0.365   $\pm$   0.026   &   0.161   $\pm$   0.050   &   0.504   &   0.243   &   0.147   &   33.0    &   18.0    \\
KIG 716 &   122 &   b   &   0.090   $\pm$   0.031   &               &   0.090   $\pm$   0.031   &       &       &       &       &       \\
KIG 719 &   98  &   b   &   0.450   $\pm$   0.042   &   0.423   $\pm$   0.005   &   0.209   $\pm$   0.063   &   0.673   &   0.403   &   0.237   &   13.8    &   10.5    \\
KIG 731 &   108 &   a   &   0.384   $\pm$   0.077   &   0.380   $\pm$   0.030   &   0.101   $\pm$   0.025   &   0.417   &   0.237   &   0.103   &   7.0 &   5.0 \\
KIG 743 &   24  &   c   &   0.385   $\pm$   0.040   &   0.380   $\pm$   0.066   &   0.067   $\pm$   0.017   &   0.515   &   0.220   &   0.106   &   13.0    &   11.0    \\
KIG 757 &   39  &   c   &   0.152   $\pm$   0.030   &               &   0.152   $\pm$   0.030   &       &       &       &       &       \\
KIG 795 &   92  &   b   &   0.339   $\pm$   0.088   &   0.333   $\pm$   0.018   &   0.292   $\pm$   0.058   &   0.400   &   0.091   &   0.068   &   11.0    &   6.0 \\
KIG 805 &   35  &   b   &   0.113   $\pm$   0.015   &               &   0.113   $\pm$   0.015   &       &       &       &       &       \\
KIG 807 &   25  &   a   &   0.161   $\pm$   0.074   &               &   0.161   $\pm$   0.074   &       &       &       &       &       \\
KIG 839 &   164 &   b   &   0.197   $\pm$   0.059   &               &   0.197   $\pm$   0.059   &       &       &       &       &       \\
KIG 892 &   122 &   a   &   0.208   $\pm$   0.084   &               &   0.208   $\pm$   0.084   &       &       &       &       &       \\
KIG 912 &   91  &   b   &   0.117   $\pm$   0.047   &               &   0.117   $\pm$   0.047   &       &       &       &       &       \\
KIG 924 &   165 &   a   &   0.093   $\pm$   0.018   &               &   0.093   $\pm$   0.018   &       &       &       &       &       \\
KIG 928 &   153 &   a   &   0.061   $\pm$   0.038   &               &   0.061   $\pm$   0.038   &       &       &       &       &       \\
KIG 931 &   111 &   a   &   0.284   $\pm$   0.056   &   0.200   $\pm$   0.005   &   0.246   $\pm$   0.053   &   0.301   &       &       &   5.5 &   2.0 \\
KIG 932 &   5   &   c   &   0.199   $\pm$   0.047   &   0.188   $\pm$   0.022   &   0.073   $\pm$   0.019   &   0.326   &   0.109   &   0.046   &   12.2    &   8.0 \\
KIG 943 &   178 &   b   &   0.278   $\pm$   0.039   &   0.166   $\pm$   0.014   &   0.213   $\pm$   0.044   &   0.614   &   0.272   &   0.122   &   8.0 &   5.0 \\

\hline

\end{tabular}

\begin{minipage}{16.3cm}

Column (1): KIG Name. Column (2): mean orientation angle (measured
as explained in \S\S\S~3.1.1). Column (3): bulge deprojection method
used (see \S\S\S~3.1.1 for more details). Column (4): total strength
Q$_{g}$ $\pm$ SD (standard deviation). Column (5): bar strength
Q$_{b}$ $\pm$ SD. Column (6): spiral strength Q$_{s}$ $\pm$ SD.
Column (7): A$_{2b}$. Column (8): A$_{4b}$. Column (9): A$_{6b}$.
Column (10): Fourier bar length in arcsec. Column (11): radius of
maximal bar torque r(Q$_{b}$) in arcsec.

\end{minipage}

\end{table}

\clearpage
\begin{table}

\setcounter{table}{1}
\begin{minipage}{70mm}
\caption{Mean/Median for Strength Parameters of \textbf{All}
Galaxies in our sample}
\end{minipage}
%\tablenum{4}
\begin{tabular}{lcc}
\hline\hline

Type (N) & Q$_{s}$ & Q$_{g}$ \\
     & mean$\pm$SE median & mean$\pm$SE median \\
\hline
Sb (25)    & 0.151$\pm$0.012~~0.161 & 0.285$\pm$0.019~~0.278 \\
Sbc (33)   & 0.157$\pm$0.012~~0.151 & 0.178$\pm$0.014~~0.170 \\
Sc (35)    & 0.188$\pm$0.013~~0.171 & 0.221$\pm$0.018~~0.183 \\
Sb-Sc (93) & 0.167$\pm$0.007~~0.161 & 0.223$\pm$0.012~~0.202 \\
\hline
\end{tabular}

\begin{minipage}{70mm}

Column (1): galaxy name. Column (2): spiral arm strength. Column
(3): total strength.

Note: N=number of galaxies; \emph{SE} is standard deviation of the
mean.
\end{minipage}

\end{table}

\begin{table}

\begin{minipage}{155mm}
\caption{Mean/Median for Strength Parameters of \textbf{Barred}
Galaxies in our sample}
\end{minipage}

\begin{tabular}{lccccc}
\hline\hline

Type (N) & Q$_{b}$ & Q$_{s}$ & Q$_{g}$ & A$_{2b}$ & l$_{bar}$ (kpc) \\
     & mean$\pm$SE median & mean$\pm$SE median & mean$\pm$SE median & mean$\pm$SE median & mean$\pm$SE median\\
\hline
Sb (22)    & 0.261$\pm$0.022~~0.225 & 0.146$\pm$0.011~~0.152 & 0.298$\pm$0.019~~0.286 & 0.51$\pm$0.03~~0.51 & 6.41$\pm$0.60~~6.02\\
Sbc (10)   & 0.206$\pm$0.031~~0.205 & 0.164$\pm$0.026~~0.124 & 0.232$\pm$0.030~~0.232 & 0.32$\pm$0.07~~0.29 & 4.44$\pm$0.86~~4.36\\
Sc (14)    & 0.242$\pm$0.033~~0.235 & 0.199$\pm$0.013~~0.186 & 0.282$\pm$0.027~~0.273 & 0.25$\pm$0.04~~0.22 & 2.32$\pm$0.43~~2.01\\
Sb-Sc (46) & 0.243$\pm$0.015~~0.222 & 0.166$\pm$0.009~~0.165 & 0.279$\pm$0.014~~0.273 & 0.39$\pm$0.03~~0.38 & 4.74$\pm$0.44~~4.36\\
\hline
\end{tabular}

\begin{minipage}{145mm}

Column (1): galaxy name. Column (2): bar strength. Column (3):
spiral arm strength. Column (4): total strength. Column (5):
$A_{2b}=(I_{2}/I_{0})_{max}$. Column (6) length of the bar in kpc.

Note: N=number of galaxies; \emph{SE} is standard deviation of the
mean.
\end{minipage}

\end{table}

\begin{table}

\setcounter{table}{3}
\begin{minipage}{44mm}
\caption{Mean/Median for Strength Parameters of \textbf{Non-Barred}
Galaxies in our sample}
\end{minipage}
%\tablenum{4}
\begin{tabular}{lc}
\hline\hline

Type (N) & Q$_{s}$ = Q$_{g}$ \\
     & mean$\pm$SE median \\
\hline
Sb (3)     & 0.192$\pm$0.062~~0.175 \\
Sbc (23)   & 0.155$\pm$0.013~~0.152 \\
Sc (21)    & 0.181$\pm$0.020~~0.149 \\
Sb-Sc (47) & 0.169$\pm$0.011~~0.152 \\
\hline
\end{tabular}

\begin{minipage}{43mm}

Column (1): galaxy name. Column (2): spiral arm strength = total
strength.

Note: N=number of galaxies; \emph{SE} is standard deviation of the
mean. For non-barred galaxies $Q_{b} \approx 0$, therefore $Q_{g}
\simeq Q_{s}$.
\end{minipage}

\end{table}

\clearpage

\begin{table}
\begin{minipage}{70mm}
\caption{Mean/Median for Strength Parameters of \textbf{All}
Galaxies in the OSU sample}
\end{minipage}
%\tablenum{4}
\begin{tabular}{lcc}
\hline\hline

Type (N) & Q$_{s}$ & Q$_{g}$ \\
     & mean$\pm$SE median & mean$\pm$SE median \\
\hline
Sb (25)    & 0.113$\pm$0.018~~0.097 & 0.205$\pm$0.025~~0.197 \\
Sbc (32)   & 0.187$\pm$0.022~~0.157 & 0.256$\pm$0.026~~0.254 \\
Sc (35)    & 0.156$\pm$0.011~~0.145 & 0.253$\pm$0.027~~0.202 \\
Sb-Sc (92) & 0.155$\pm$0.010~~0.132 & 0.241$\pm$0.015~~0.210 \\
\hline
\end{tabular}

\begin{minipage}{70mm}

Column (1): galaxy name. Column (2): spiral arm strength. Column
(3): total strength.

Note: N=number of galaxies; \emph{SE} is standard deviation of the
mean.
\end{minipage}

\end{table}

\begin{table}

\begin{minipage}{155mm}
\caption{Mean/Median for Strength Parameters of \textbf{Barred}
Galaxies in the OSU sample}
\end{minipage}

\begin{tabular}{lccccc}
\hline\hline

Type (N) & Q$_{b}$ & Q$_{s}$ & Q$_{g}$ & A$_{2b}$ & l$_{bar}$ (kpc) \\
     & mean$\pm$SE median & mean$\pm$SE median & mean$\pm$SE median & mean$\pm$SE median & mean$\pm$SE median\\
\hline
Sb (20)    & 0.204$\pm$0.023~~0.196 & 0.111$\pm$0.014~~0.099 & 0.227$\pm$0.023~~0.227 & 0.48$\pm$0.04~~0.45 & 4.33$\pm$0.78~~3.13\\
Sbc (19)   & 0.240$\pm$0.033~~0.225 & 0.194$\pm$0.030~~0.169 & 0.310$\pm$0.034~~0.259 & 0.50$\pm$0.05~~0.42 & 3.72$\pm$0.60~~3.04\\
Sc (21)    & 0.290$\pm$0.038~~0.321 & 0.165$\pm$0.015~~0.167 & 0.318$\pm$0.038~~0.360 & 0.39$\pm$0.04~~0.35 & 2.42$\pm$0.40~~1.89\\
Sb-Sc (60) & 0.246$\pm$0.019~~0.212 & 0.156$\pm$0.013~~0.137 & 0.285$\pm$0.019~~0.254 & 0.45$\pm$0.03~~0.41 & 3.47$\pm$0.36~~2.64\\
\hline
\end{tabular}

\begin{minipage}{145mm}

Column (1): galaxy name. Column (2): bar strength. Column (3):
spiral arm strength. Column (4): total strength. Column (5):
$A_{2b}=(I_{2}/I_{0})_{max}$. Column (6) length of the bar in kpc.

Note: N=number of galaxies; \emph{SE} is standard deviation of the
mean.
\end{minipage}

\end{table}

\begin{table}

\begin{minipage}{44mm}
\caption{Mean/Median for Strength Parameters of \textbf{Non-Barred}
Galaxies in the OSU sample}
\end{minipage}
%\tablenum{4}
\begin{tabular}{lc}
\hline\hline

Type (N) & Q$_{g}$ \\
     & mean$\pm$SE median \\
\hline
Sb (5)     & 0.117$\pm$0.075~~0.039 \\
Sbc (13)   & 0.176$\pm$0.031~~0.148 \\
Sc (14)    & 0.155$\pm$0.015~~0.148 \\
Sb-Sc (32) & 0.158$\pm$0.018~~0.143 \\
\hline
\end{tabular}

\begin{minipage}{43mm}

Column (1): galaxy name. Column (2): total strength.

Note: N=number of galaxies; \emph{SE} is standard deviation of the
mean.
\end{minipage}

\end{table}

\clearpage
\begin{table}
\begin{minipage}{45mm}
\caption{Spiral Arm Multiplicities for a Selected Number of Galaxies
in our Sample (N=86)}
\end{minipage}
\begin{tabular}{lc}
\hline\hline

Multiplicity  &  Number of galaxies      \\
~~~~~~m             & N       \\
\hline
1~................ &  1  \\
2~................ & 34  \\
3~................ &  3  \\
4~................ &  2  \\
1 \& 2~.........   & 11  \\
1 \& 3~.........   &  2  \\
1 \& 4~.........   &  2  \\
2 \& 3~.........   & 24  \\
2 \& 4~.........   &  2  \\
3 \& 4~.........   &  1  \\
1, 2 \& 3~.....    &  2  \\
1, 2 \& 4~.....    &  1  \\
2, 3 \& 4~.....    &  1  \\
\hline
\end{tabular}

\begin{minipage}{45mm}

Column (1): Spiral Arm multiplicities present in our sample. Column
(2): Number of galaxies in each spiral arm multiplicity bin.

\end{minipage}

\end{table}

\begin{figure*}
\centering
\includegraphics[width=\columnwidth]{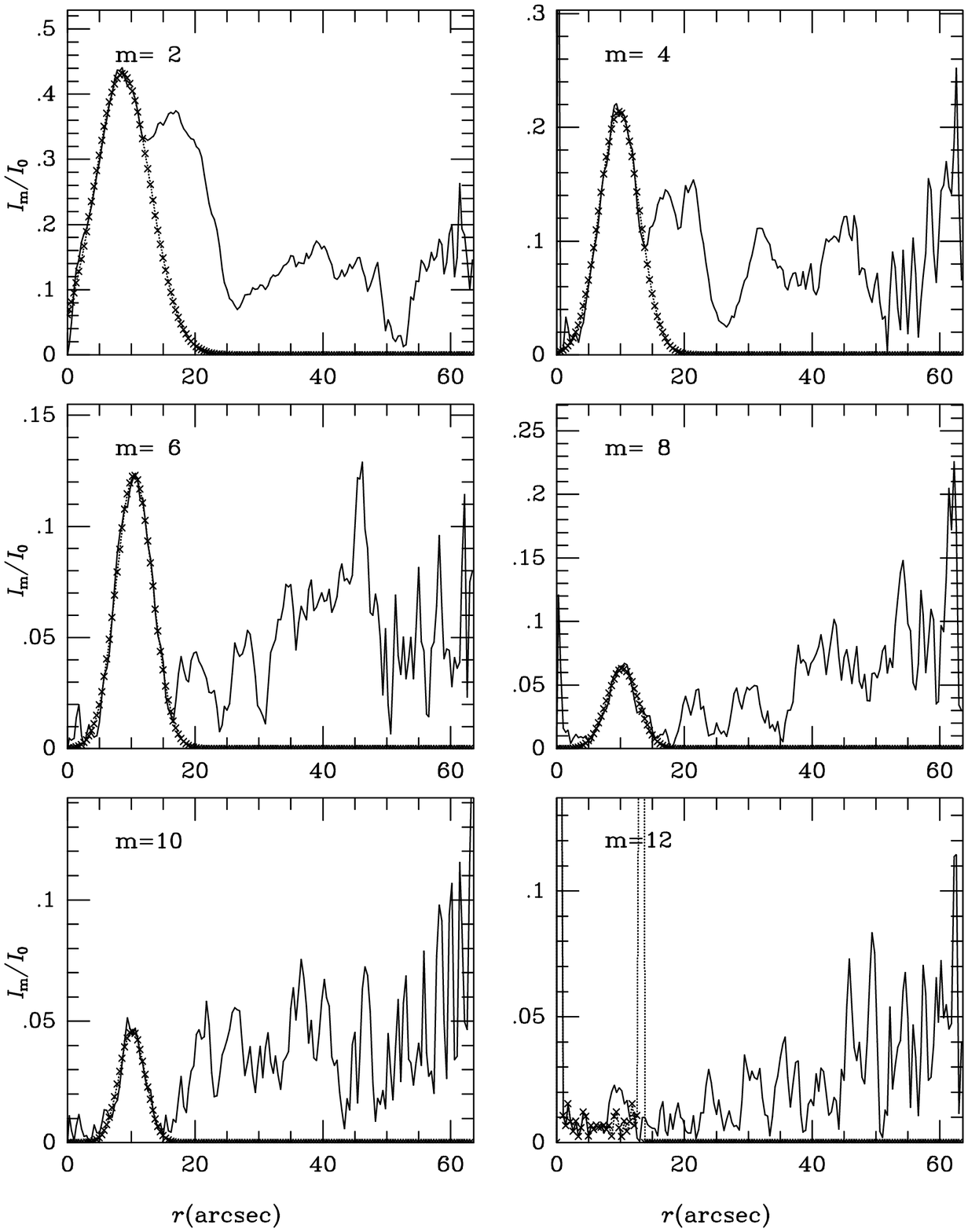}
\hspace{0.5cm}
\includegraphics[width=\columnwidth]{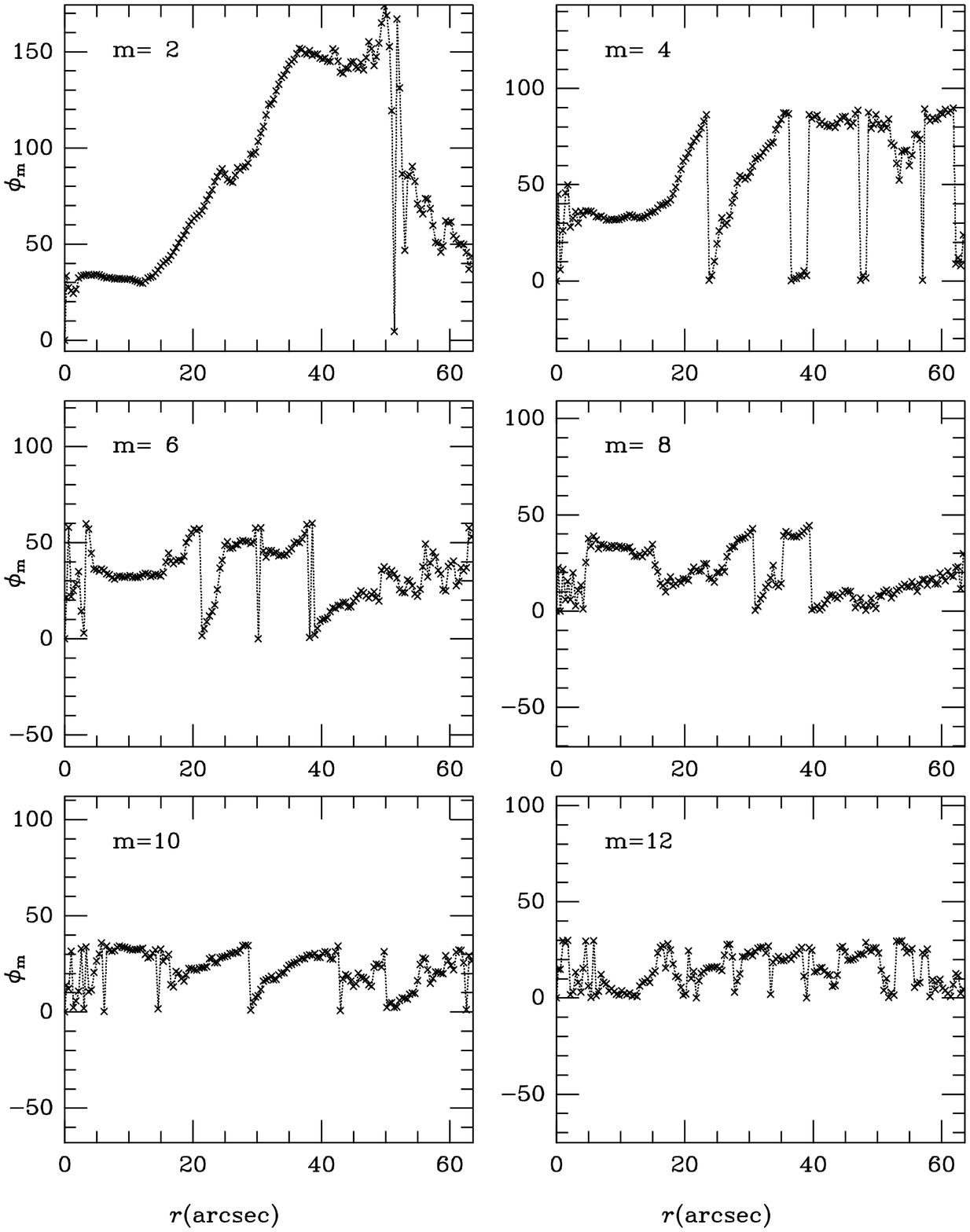}
\caption{KIG 550: ($left$): Relative Fourier intensity amplitudes
$I_{m}/I_{0}$ for the first six even Fourier terms (m=2 to m=12);
($right$): Phase profiles $\phi_{m}$ for the first six even Fourier
terms (m=2 to m=12).}
\end{figure*}
\clearpage

\begin{figure}
\plotone{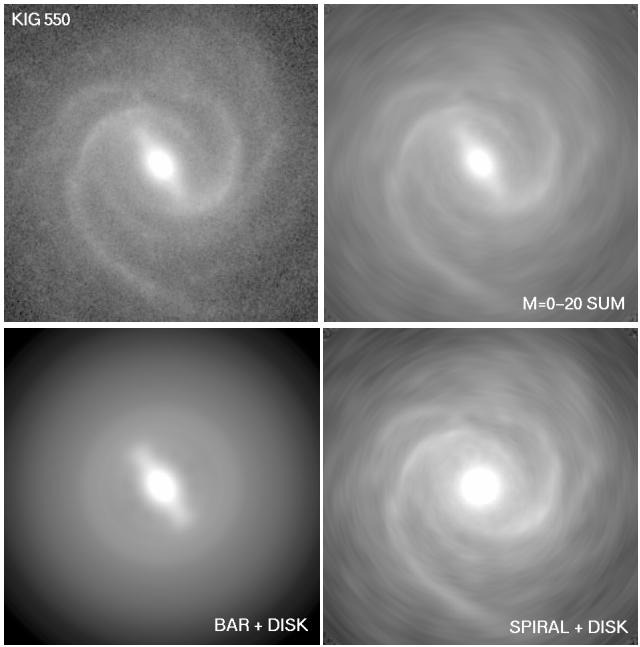} \caption{KIG 550: ($upper-left$) original
reduced/deprojected i-band image; ($upper-right$) ``M=0-20 SUM''
image (``Fourier-smoothed'' version of the original image) = the sum
of the 21 Fourier terms; ($lower-left$) ``BAR + DISK'' image = the
sum of the bar image and m=0 image; ($lower-right$) ``SPIRAL +
DISK'' image = ``M=0-20 SUM'' image minus the bar image. }
\label{fig2}
\end{figure}

\clearpage

\begin{figure*}
\centering
\includegraphics[width=\columnwidth]{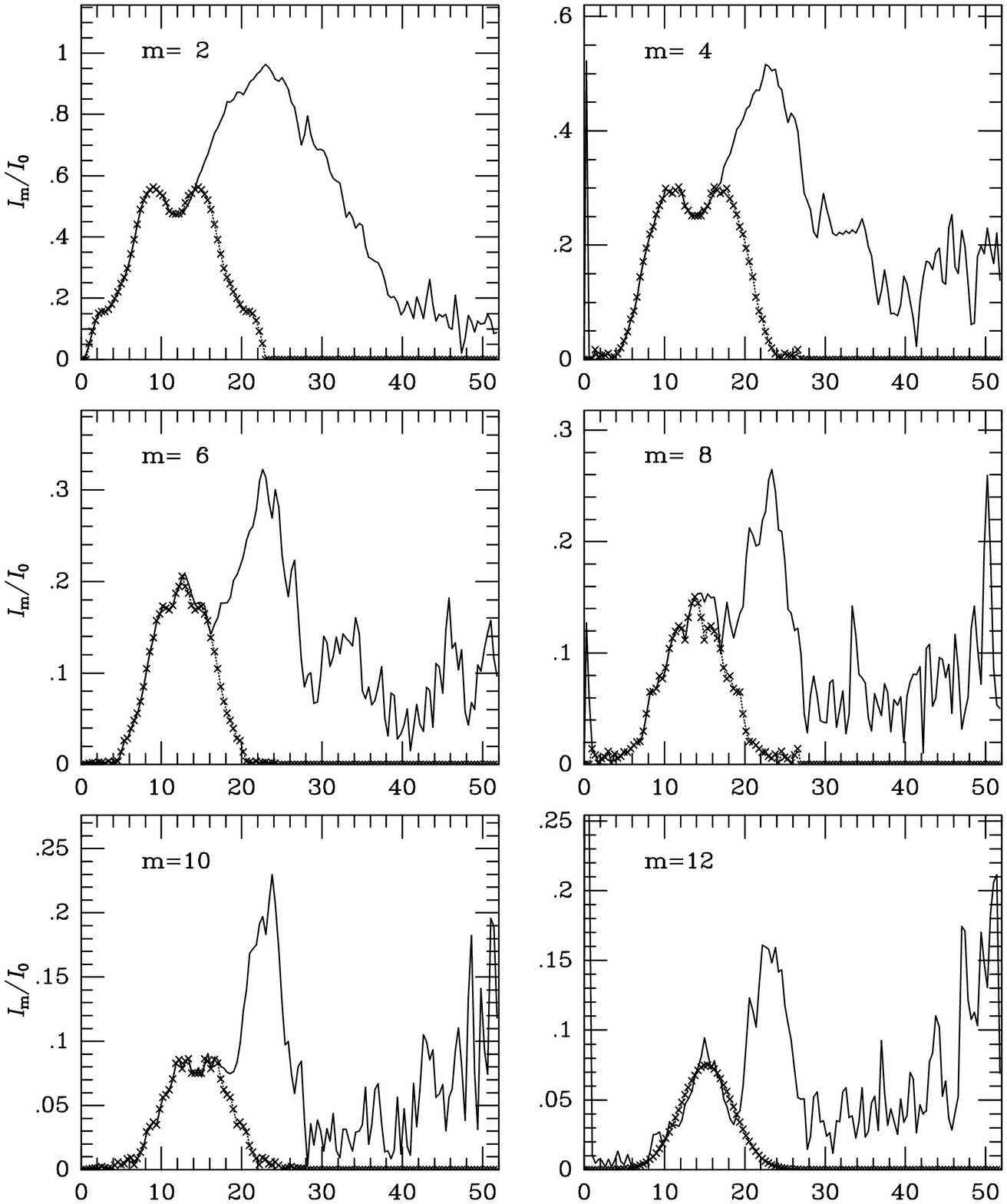}
\hspace{0.5cm}
\includegraphics[width=\columnwidth]{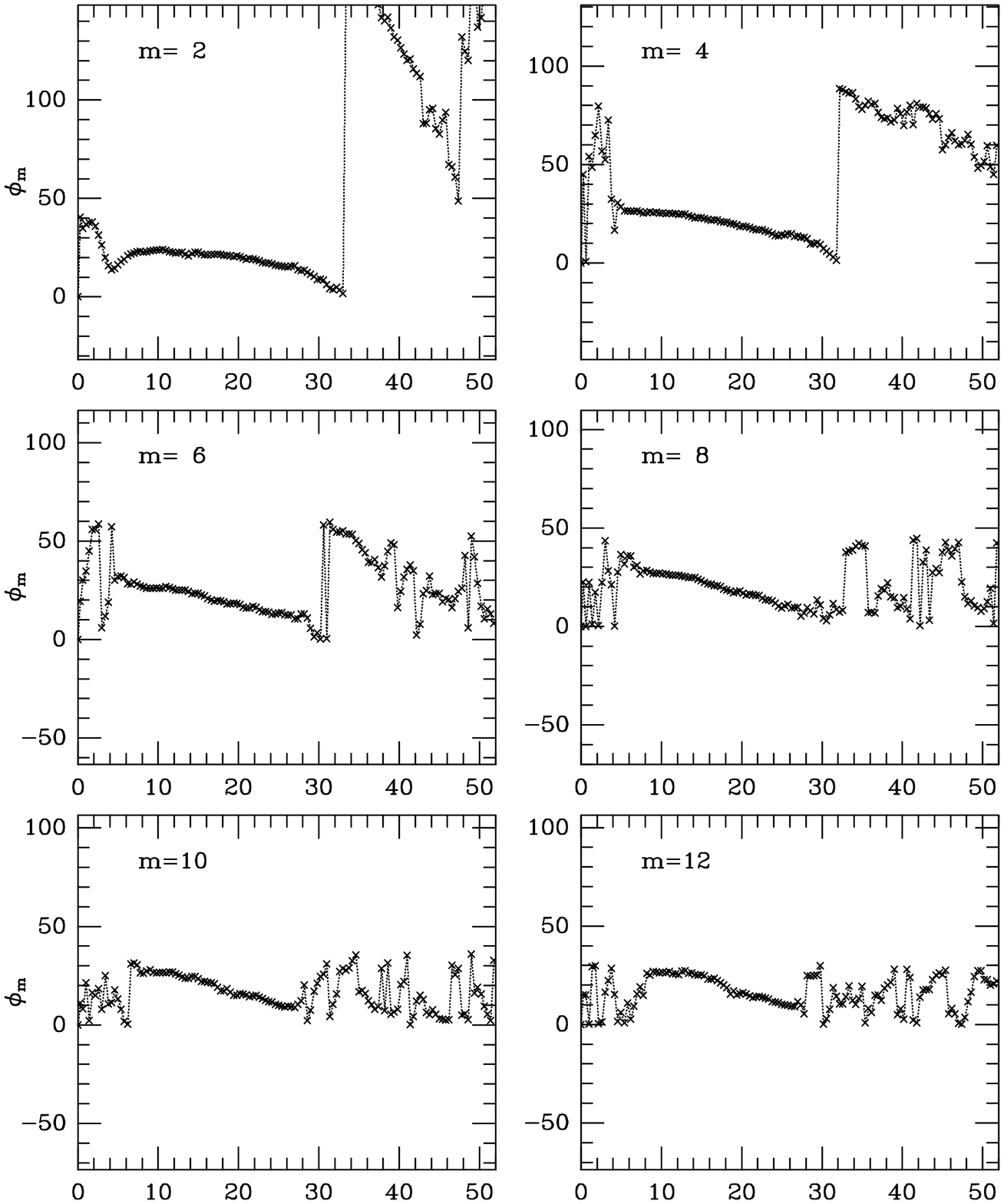}
\includegraphics[width=\columnwidth]{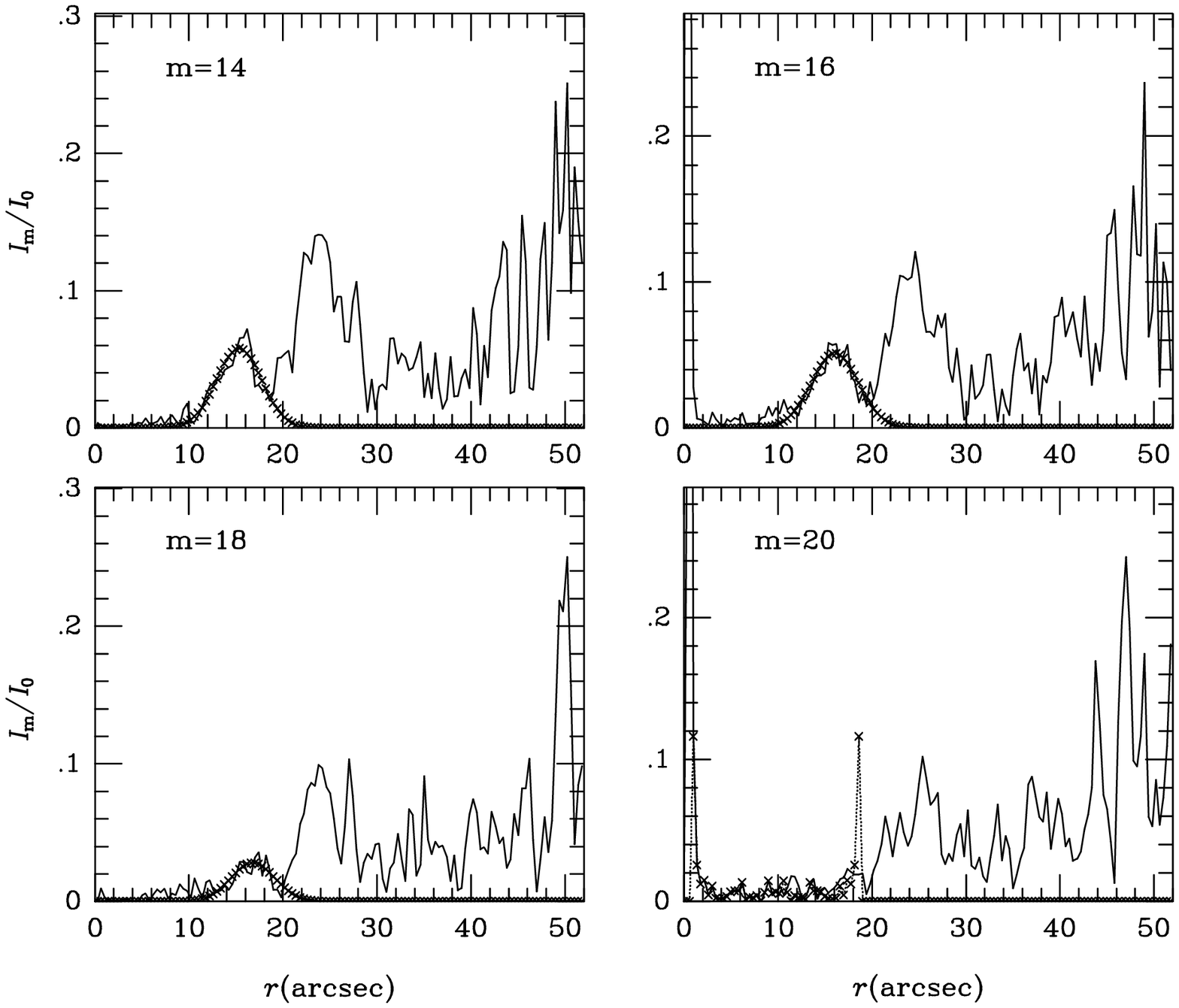}
\hspace{0.5cm}
\includegraphics[width=\columnwidth]{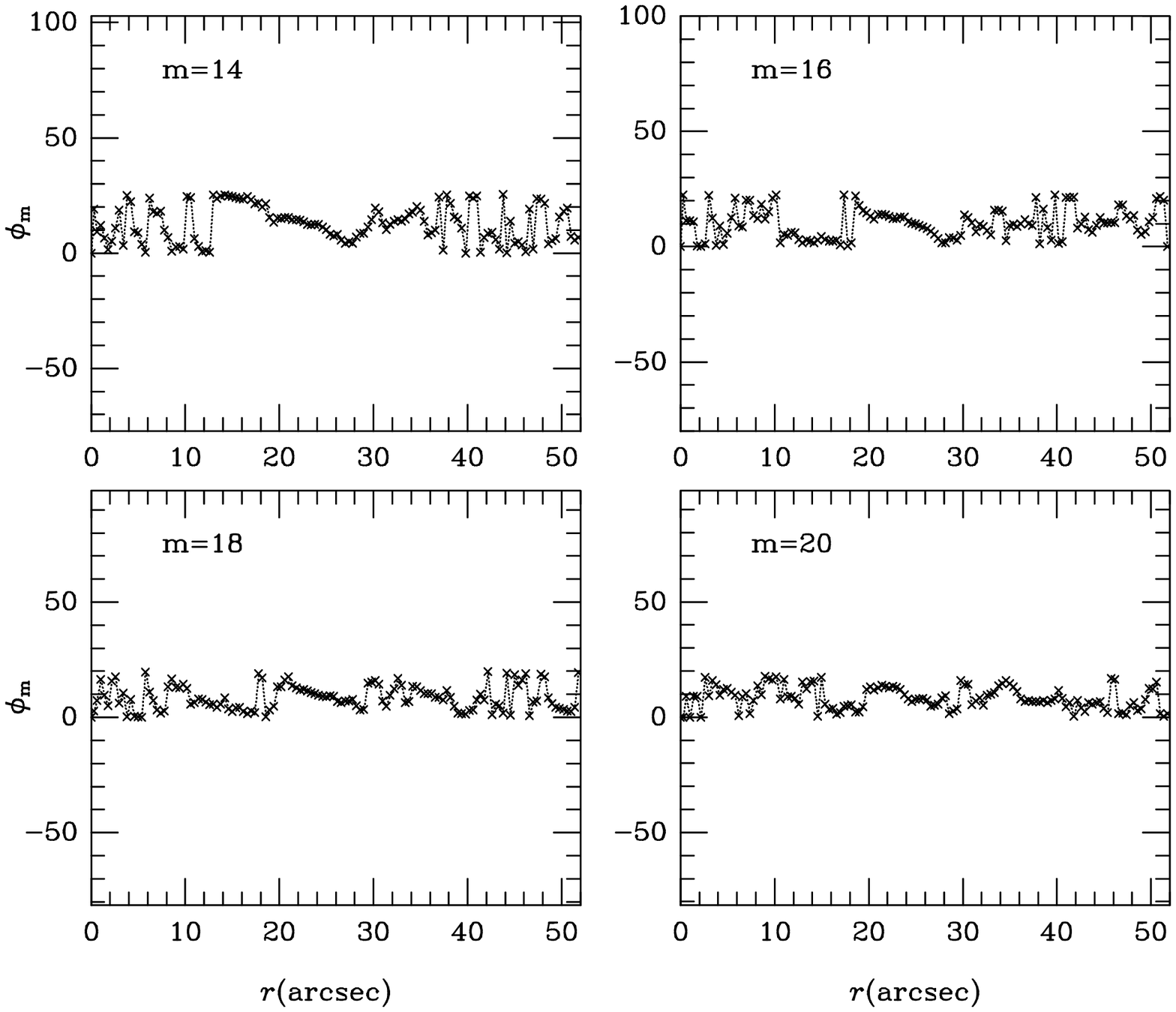}
\caption{KIG 553: ($left$): Relative Fourier intensity amplitudes
$I_{m}/I_{0}$ for the first ten even Fourier terms (m=2 to m=20);
($right$): Phase profiles $\phi_{m}$ for the first ten even Fourier
terms (m=2 to m=20).}
\end{figure*}

\clearpage

\begin{figure}
\plotone{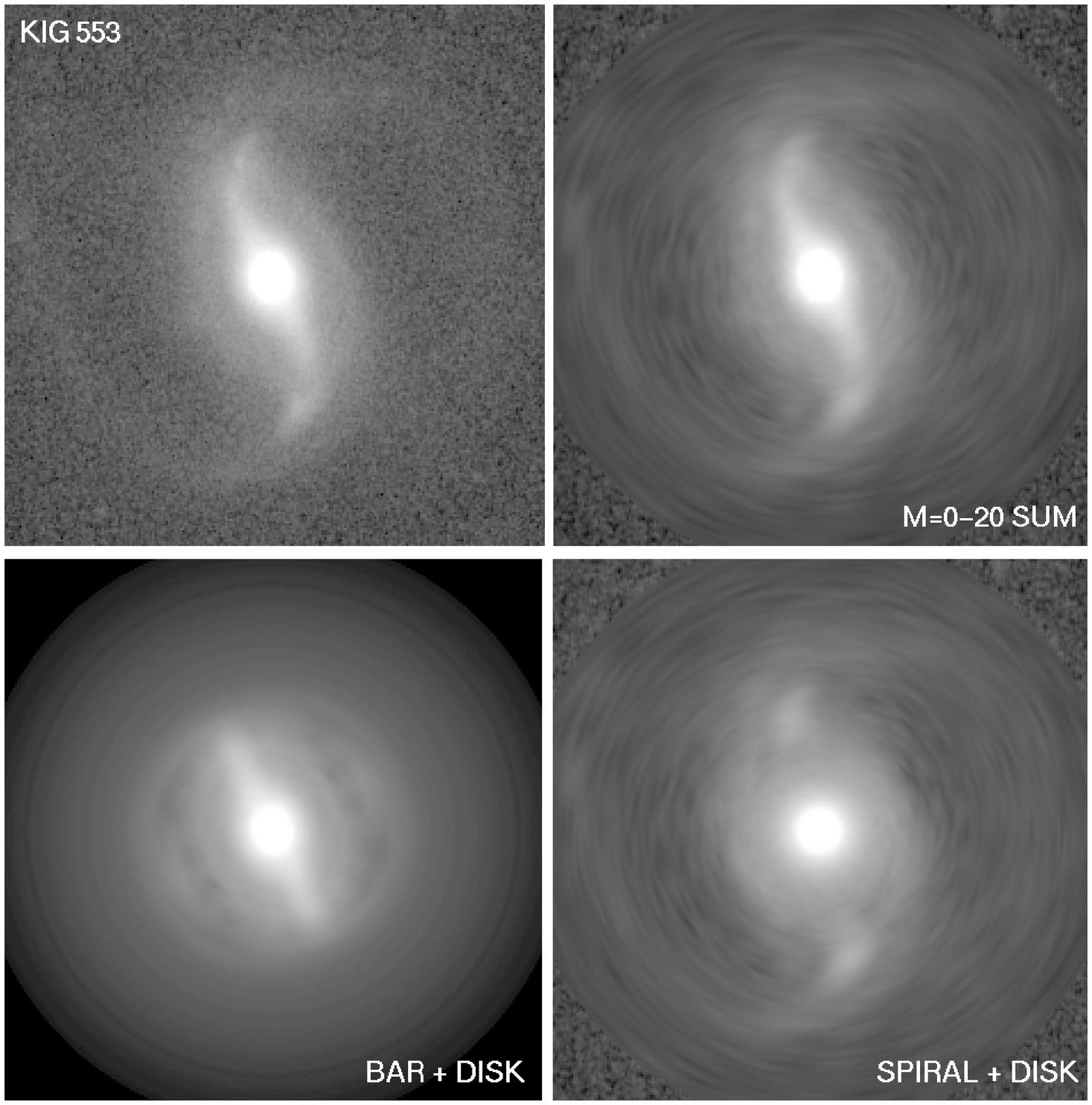} \caption{KIG 553: The designation of each image
is the same as in Figure 2.} \label{fig4}
\end{figure}

\clearpage

\begin{figure*}
\centering
\includegraphics[width=\columnwidth]{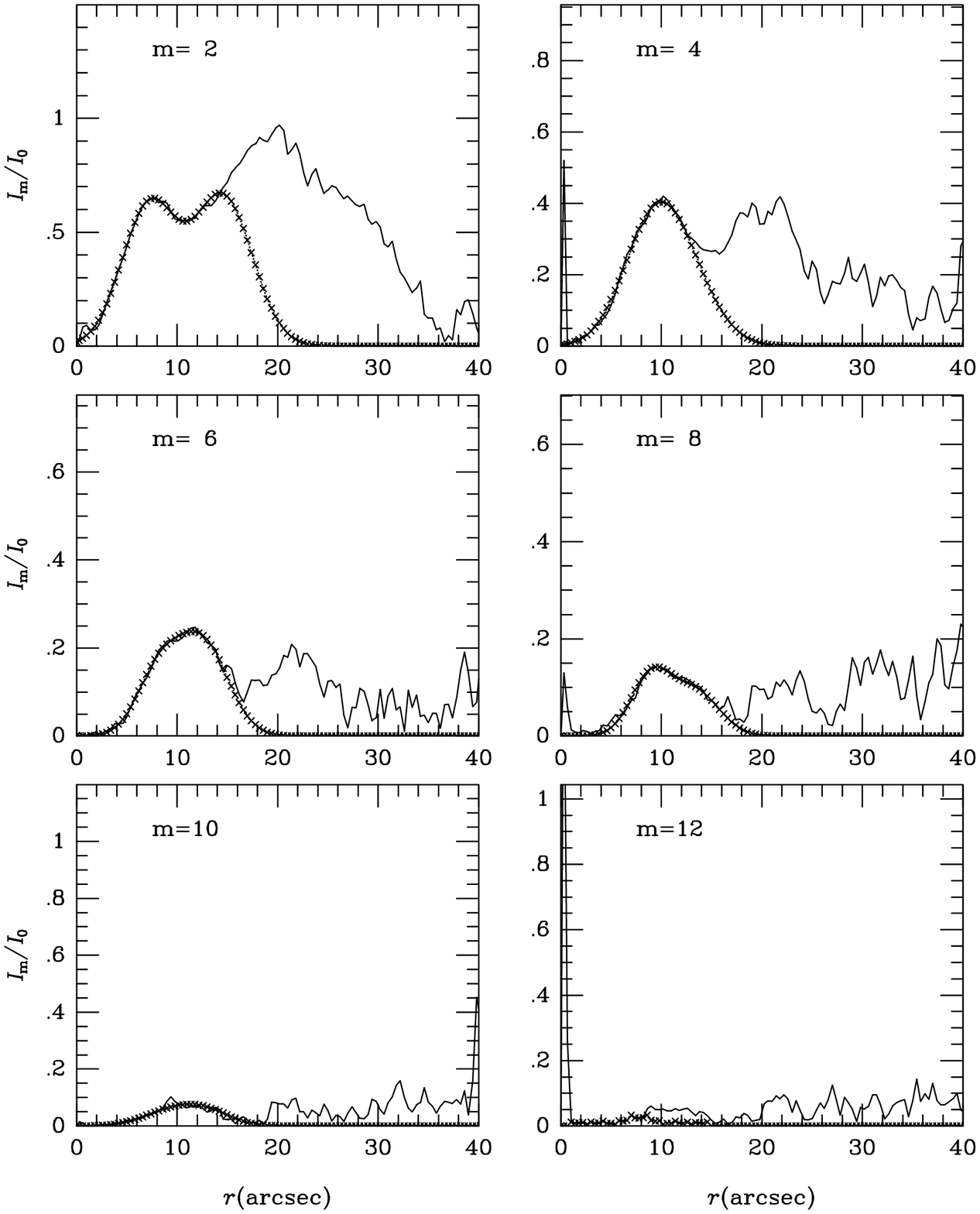}
\hspace{0.5cm}
\includegraphics[width=\columnwidth]{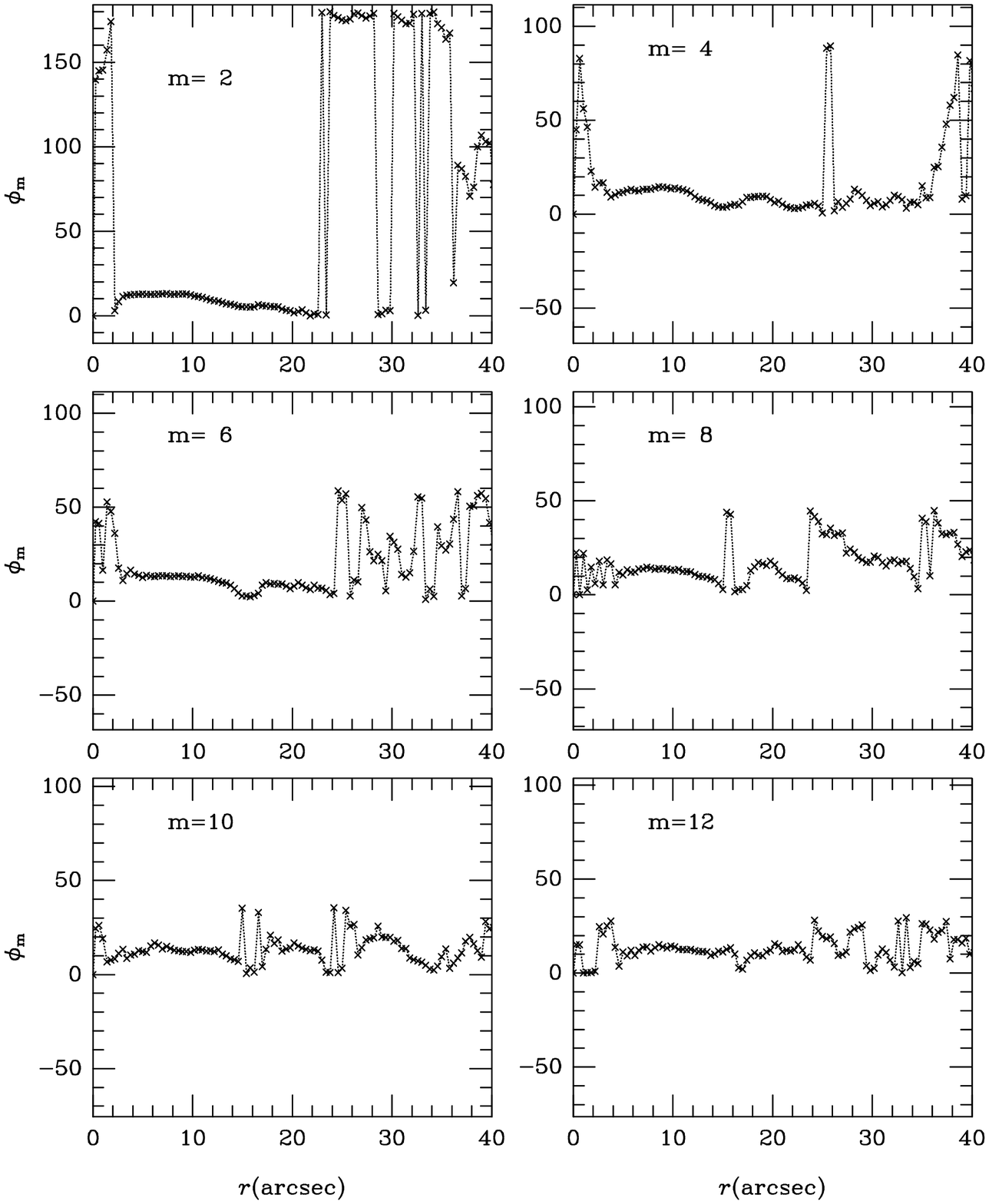}
\caption{KIG 719: ($left$): Relative Fourier intensity amplitudes
$I_{m}/I_{0}$ for the first six even Fourier terms (m=2 to m=12);
($right$): Phase profiles $\phi_{m}$ for the first six even Fourier
terms (m=2 to m=12).}
\end{figure*}
\clearpage

\begin{figure}
\plotone{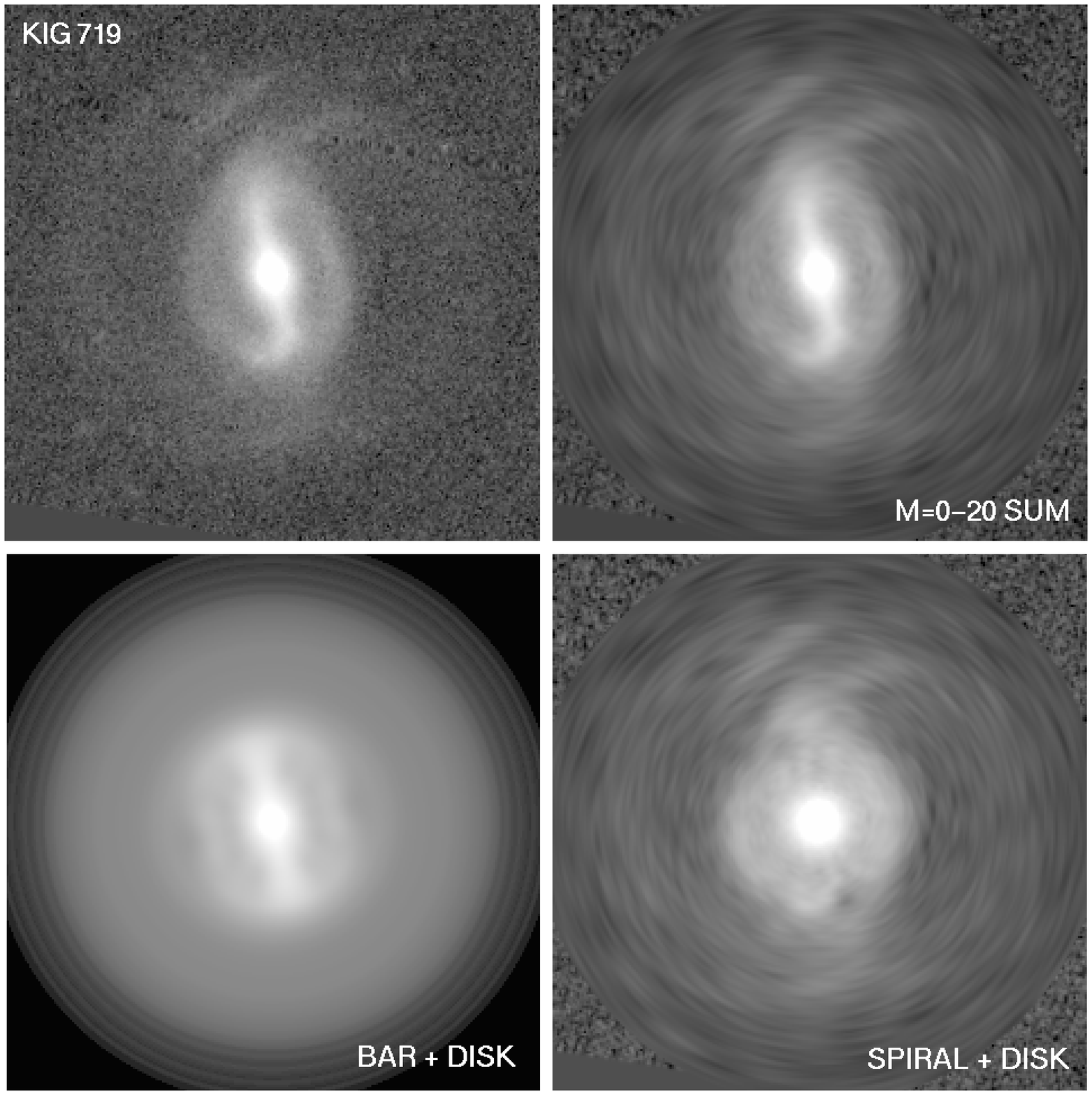} \caption{KIG 719: The designation of each image
is the same as in Figure 2.} \label{fig6}
\end{figure}

\clearpage

\begin{figure}
\plotone{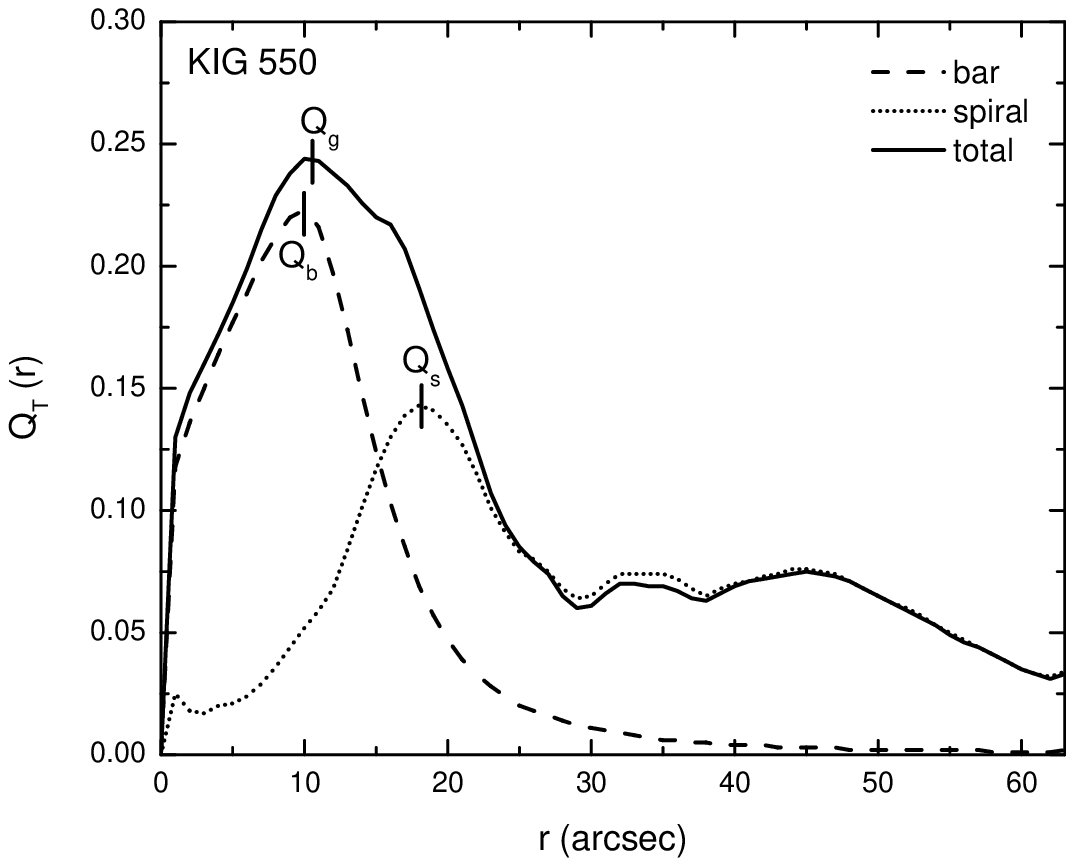} \caption{KIG 550: The relative strength of the
perturbation $Q_{T}(r)$ as a function of radius for the bar (dashed
line), spiral structure (dotted line) and total (solid line). Bar
strength ($Q_{b}$), spiral strength ($Q_{s}$) and total strength
($Q_{g}$) are indicated on the figure.} \label{fig7}
\end{figure}

\begin{figure}
\plotone{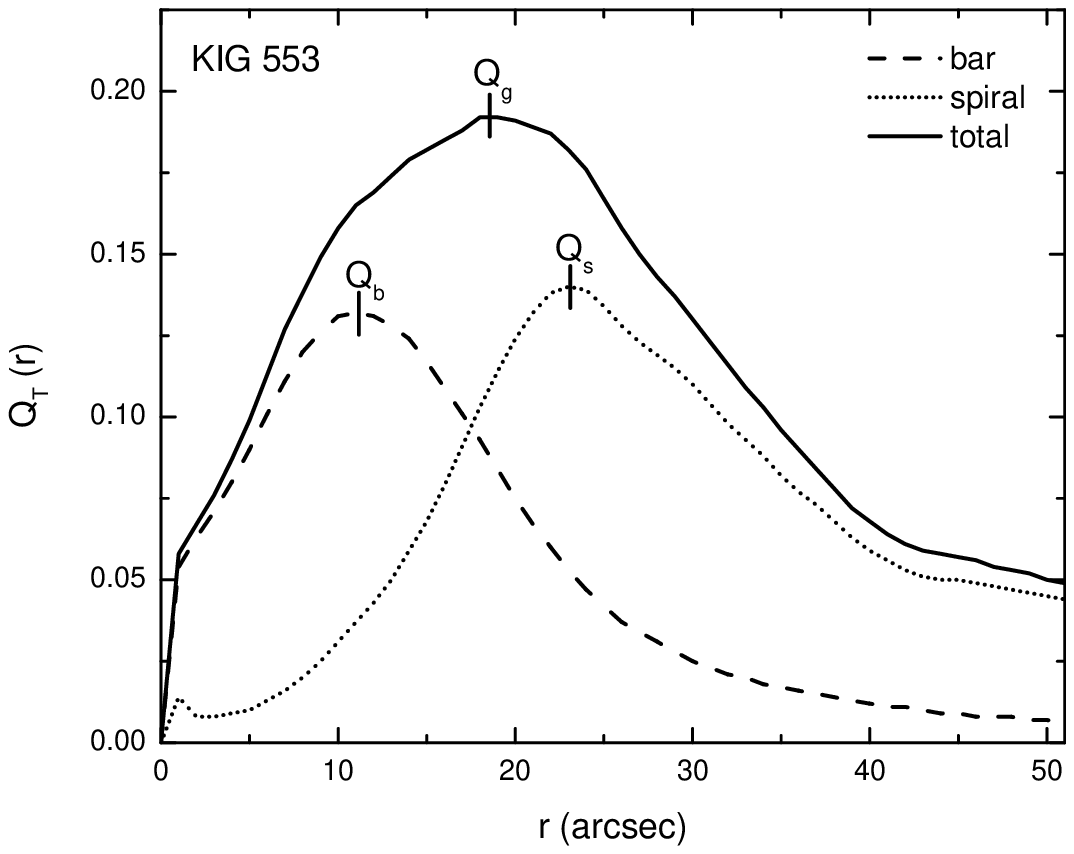} \caption{KIG 553: The relative strength of the
perturbation $Q_{T}(r)$ as a function of radius for the bar (dashed
line), spiral structure (dotted line) and total (solid line). Bar
strength ($Q_{b}$), spiral strength ($Q_{s}$) and total strength
($Q_{g}$) are indicated on the figure.} \label{fig8}
\end{figure}

\begin{figure}
\plotone{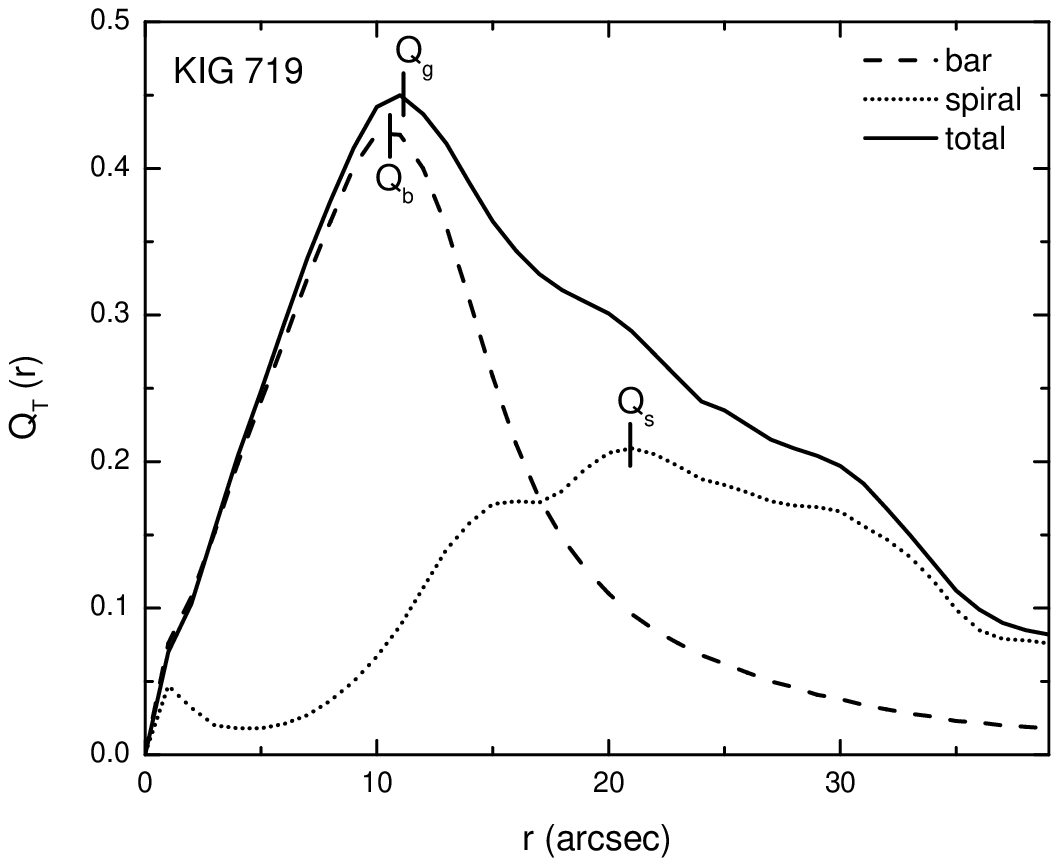} \caption{KIG 719: The relative strength of the
perturbation $Q_{T}(r)$ as a function of radius for the bar (dashed
line), spiral structure (dotted line) and total (solid line). Bar
strength ($Q_{b}$), spiral strength ($Q_{s}$) and total strength
($Q_{g}$) are indicated on the figure.} \label{fig9}
\end{figure}

\begin{figure}
\includegraphics[width=1.00\columnwidth,clip=true]{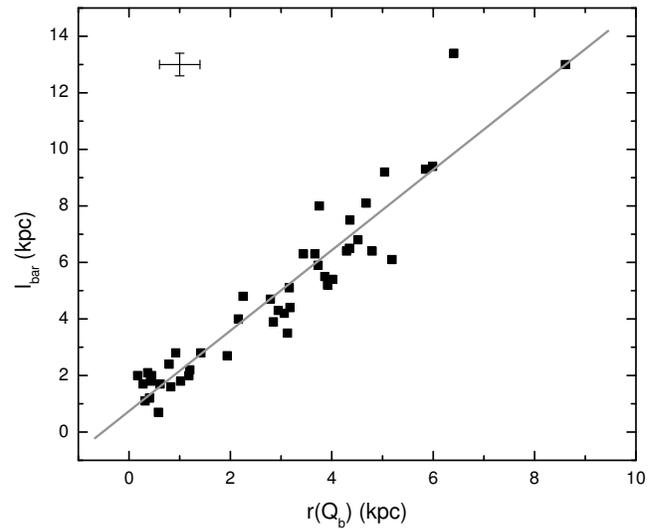}
\caption{The correlation between the bar maximal torque radius
r(Q$_{b}$) and the Fourier bar length l$_{bar}$ for the CIG/KIG
barred Sb-Sc galaxies in our sample (N=46). A linear regression fit
of slope 1.42 is shown (correlation coefficient 0.95). The 2$\sigma$
typical error bars are shown as well.} \label{fig10}
\end{figure}

\clearpage

\begin{figure}
\includegraphics[width=0.95\columnwidth,clip=true]{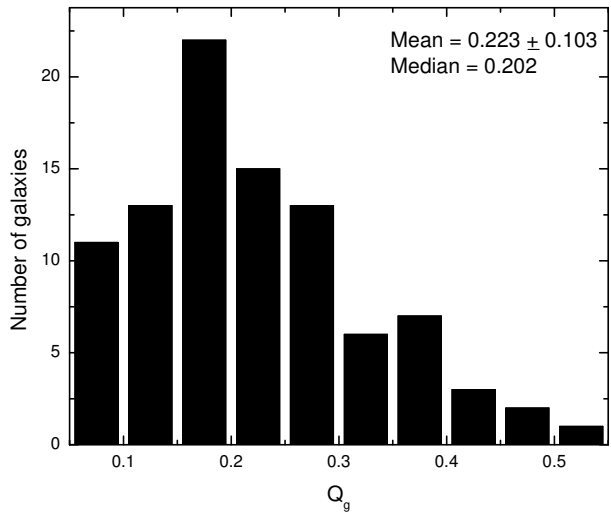}
\caption{Distribution of the total strength Q$_{g}$ for the CIG/KIG
Sb-Sc galaxies in our sample (N=93).} \label{fig11}
\end{figure}

\begin{figure}
\includegraphics[width=0.95\columnwidth,clip=true]{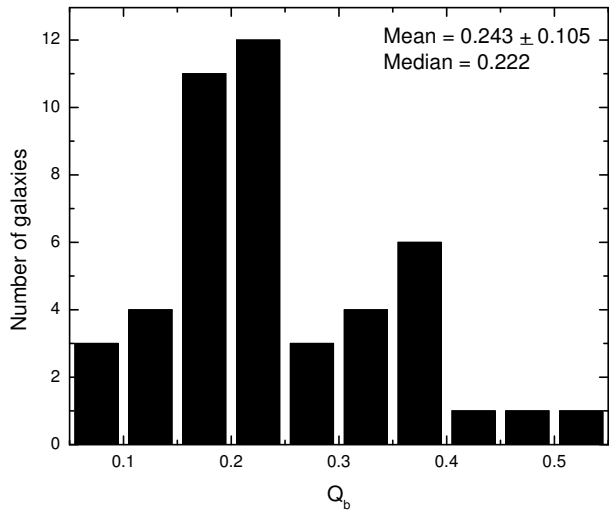}
\caption{Distribution of the bar strength Q$_{b}$ for the barred
CIG/KIG Sb-Sc galaxies in our sample (N=46).} \label{fig12}
\end{figure}

\begin{figure*}
\centering
\includegraphics[width=0.99\columnwidth]{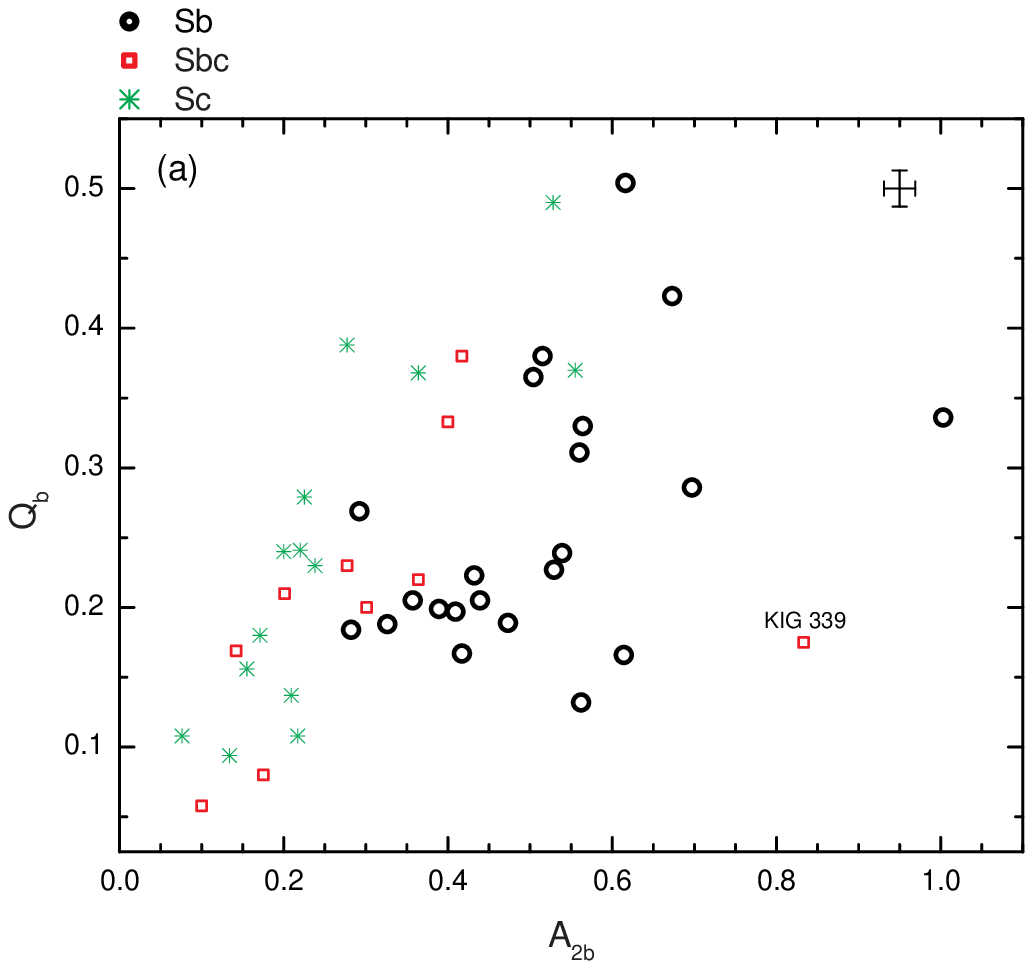}
\includegraphics[width=\columnwidth]{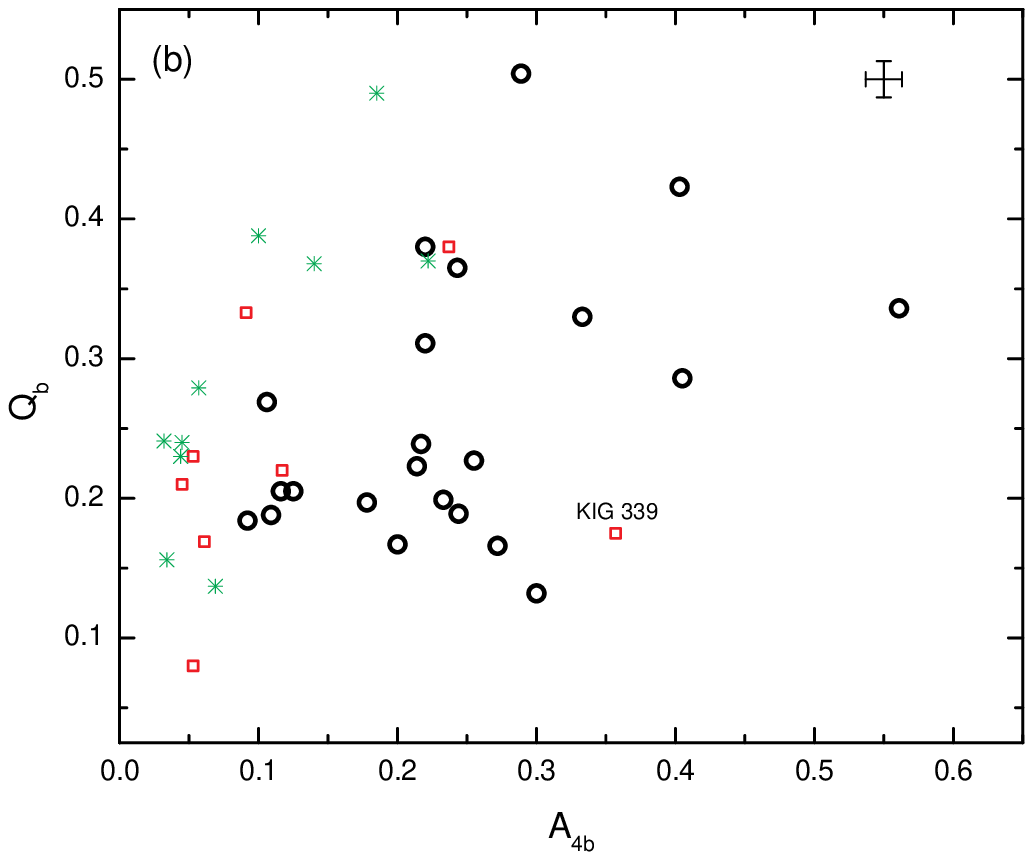}
\includegraphics[width=\columnwidth]{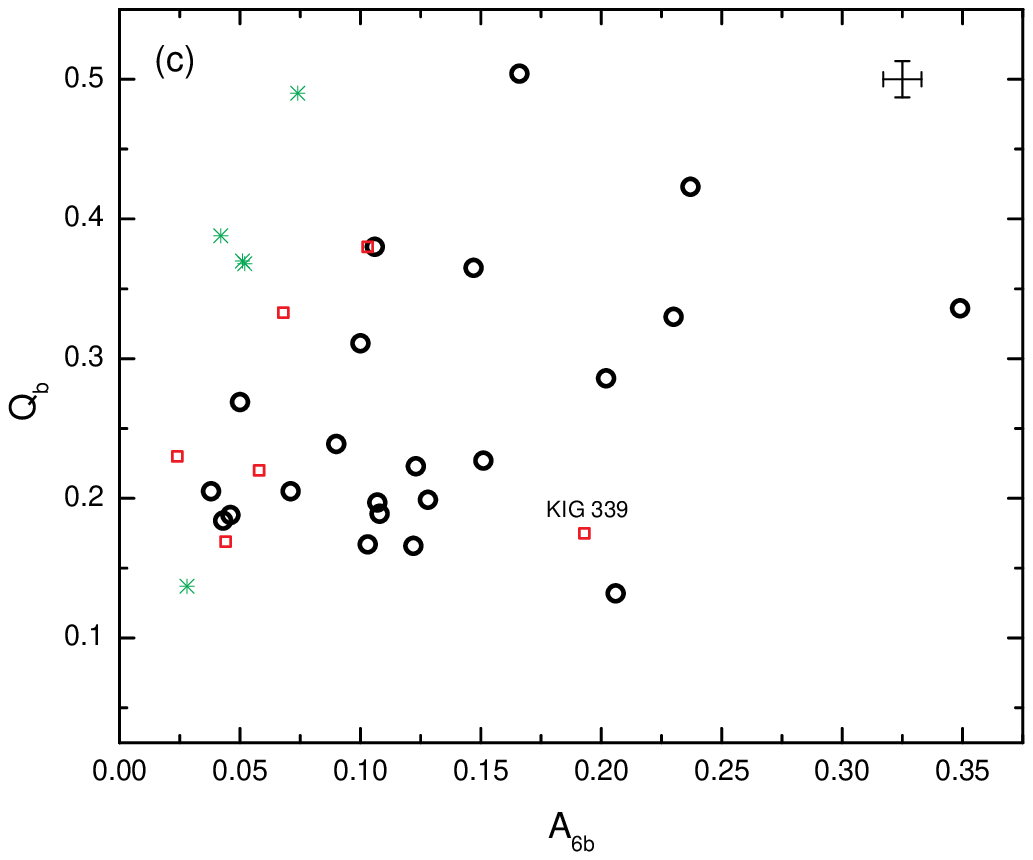}
\caption{Barred CIG/KIG Sb-Sc galaxies: (a) Bar strength Q$_{b}$
versus maximum relative Fourier intensity amplitudes at m=2,
A$_{2b}$ (N=46 galaxies); (b) Bar strength Q$_{b}$ versus maximum
relative Fourier intensity amplitudes at m=4, A$_{4b}$ (N=40
galaxies); (c) Bar strength Q$_{b}$ versus maximum relative Fourier
intensity amplitudes at m=6, A$_{6b}$ (N=33 galaxies). An outlier
(KIG 339) is labeled on the plots. Typical 2$\sigma$ error bars are
shown in each panel.}
\end{figure*}
\clearpage

\begin{figure}
\includegraphics[width=0.95\columnwidth,clip=true]{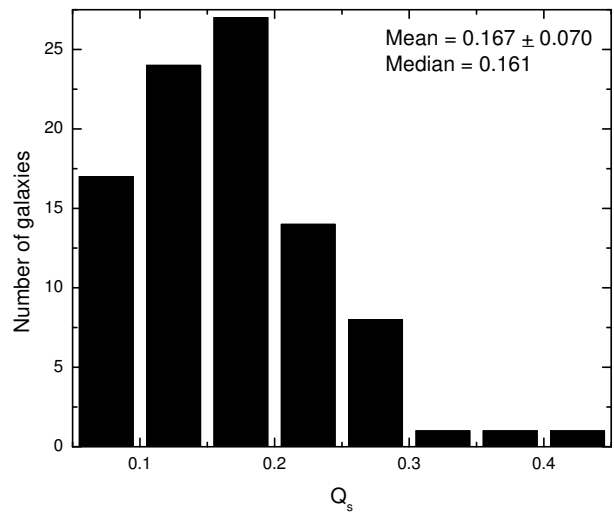}
\caption{Distribution of the spiral strengths Q$_{s}$ for the
CIG/KIG Sb-Sc galaxies in our sample (N=93).} \label{fig14}
\end{figure}

\begin{figure*}
\centering
\includegraphics[width=0.98\columnwidth,clip=true]{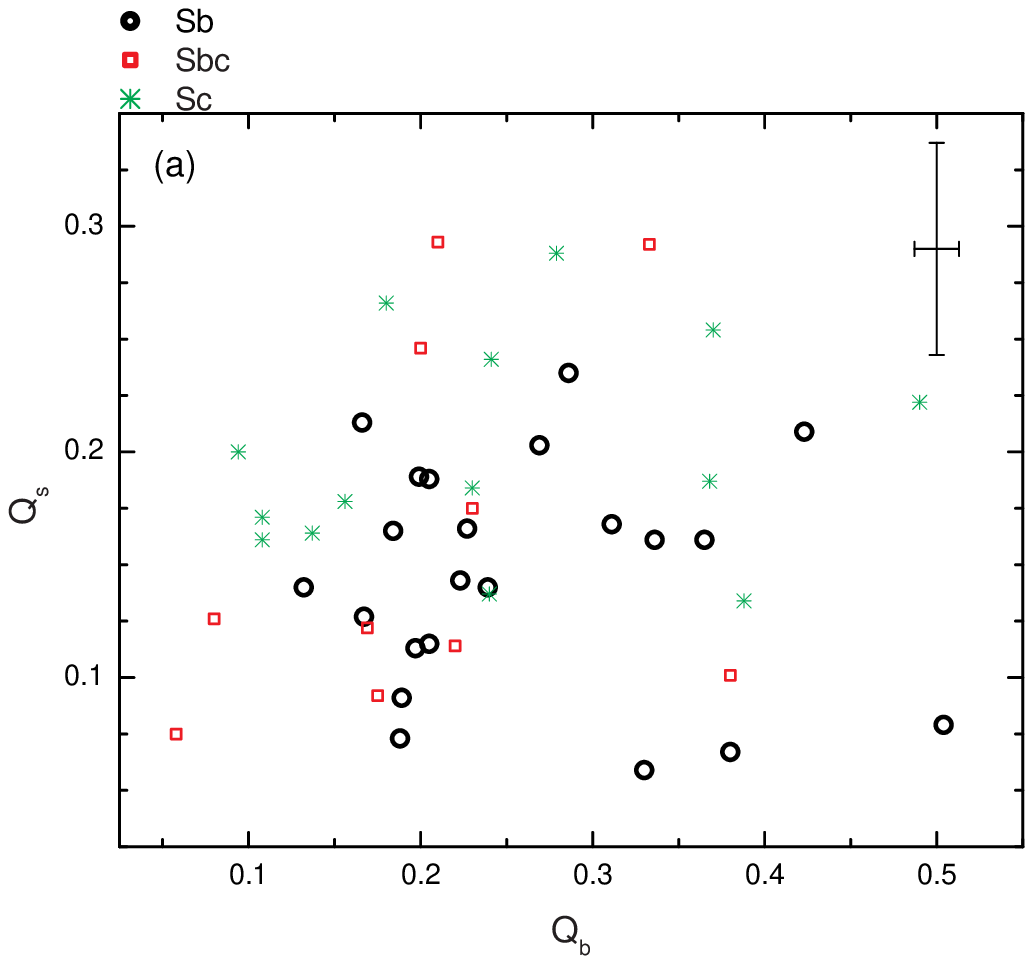}
\hspace{0.5cm}
\includegraphics[width=\columnwidth,clip=true]{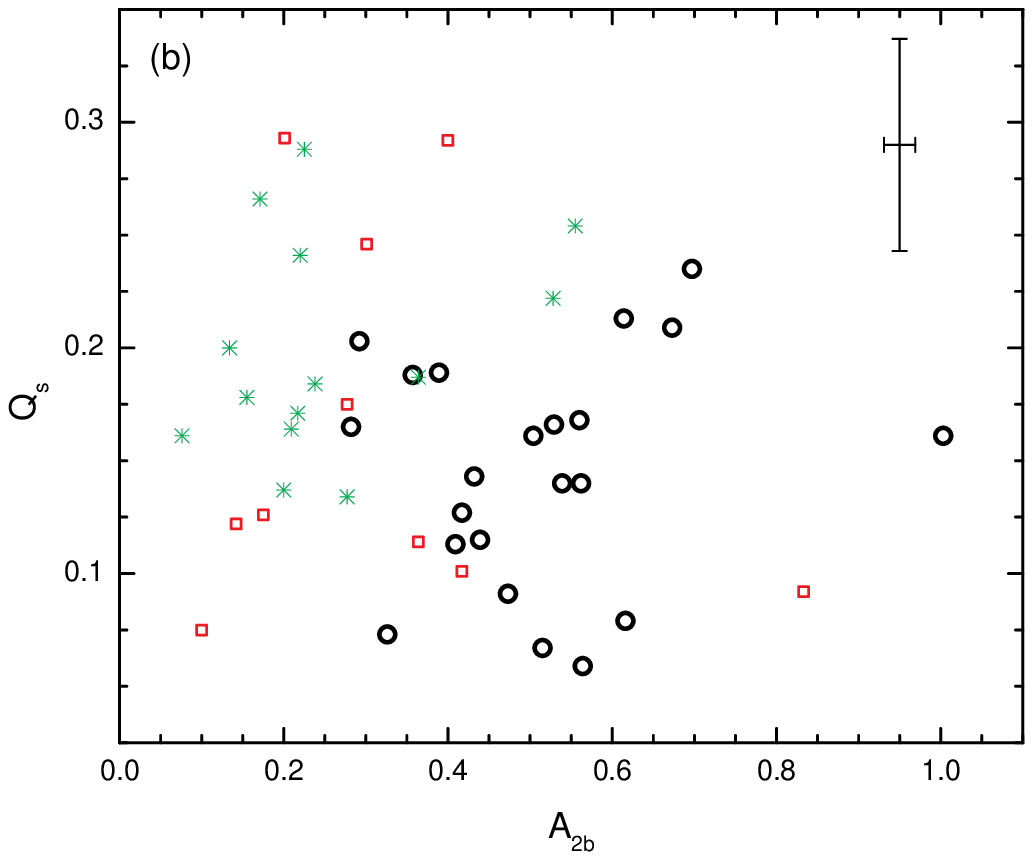}
\caption{Barred CIG/KIG Sb-Sc galaxies (N=46): (a) Spiral arm
strength Q$_{s}$ versus bar strength Q$_{b}$; (b) Spiral arm
strength Q$_{s}$ versus maximum of the relative Fourier intensity
amplitudes at m=2, A$_{2b}$. Typical 2$\sigma$ error bars are shown
in each panel.}
\end{figure*}
%\clearpage

\begin{figure*}
\includegraphics[width=0.97\columnwidth,clip=true]{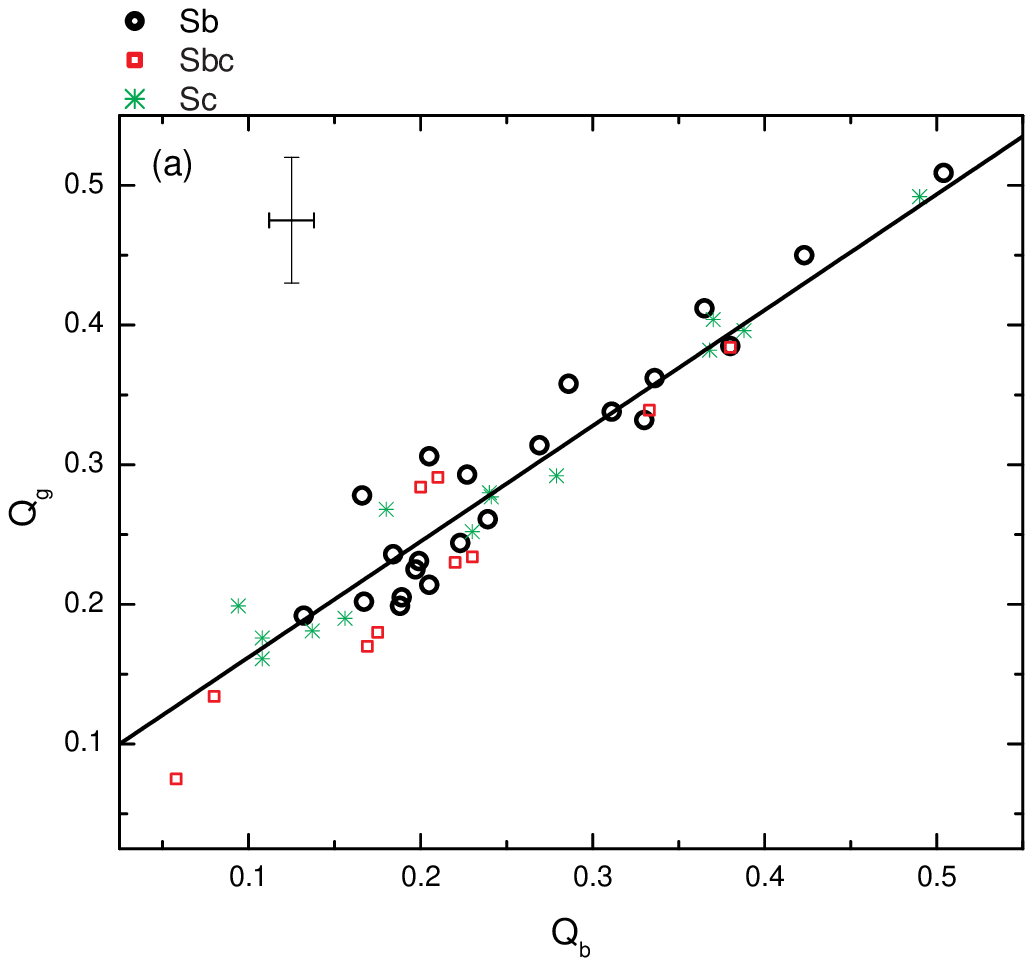}
\hspace{0.5cm}
\includegraphics[width=\columnwidth,clip=true]{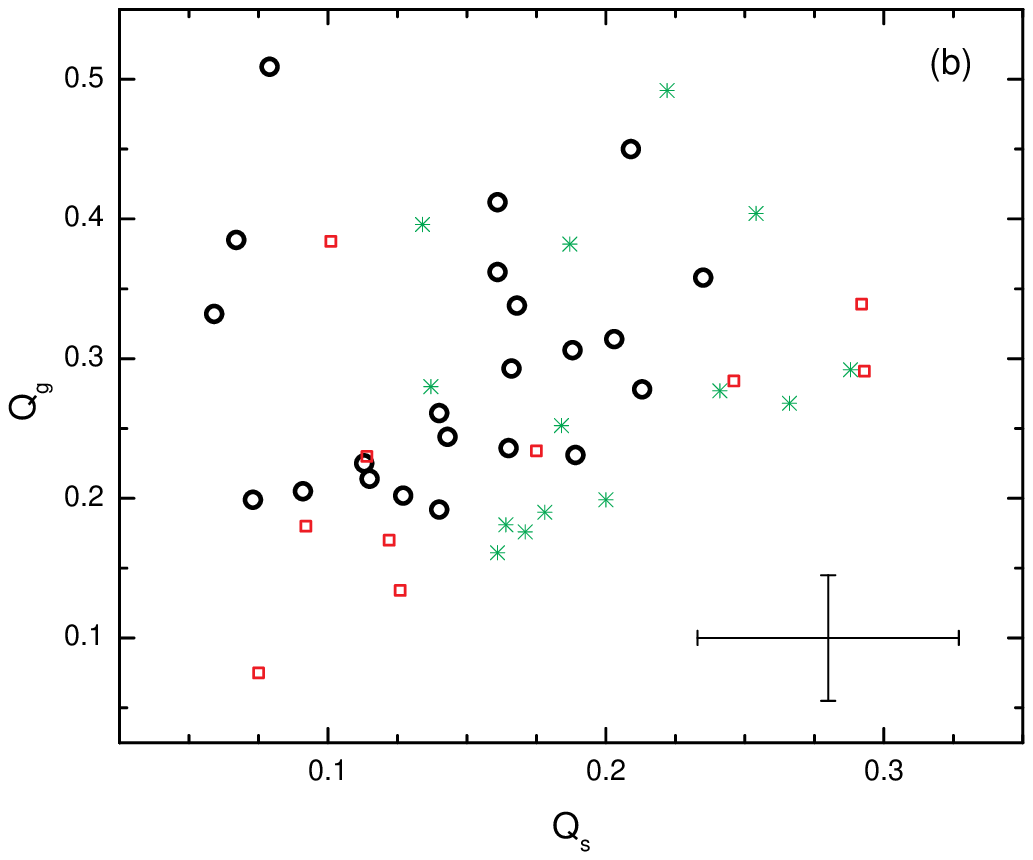}
\caption{Barred CIG/KIG Sb-Sc galaxies (N=46): (a) Total strength
Q$_{g}$ versus bar strength Q$_{b}$; (b) Total strength Q$_{g}$
versus spiral arm strength Q$_{s}$. Typical 2$\sigma$ error bars are
shown in each panel.} \label{fig16}
\end{figure*}
\clearpage

\begin{figure}
\includegraphics[width=\columnwidth,clip=true]{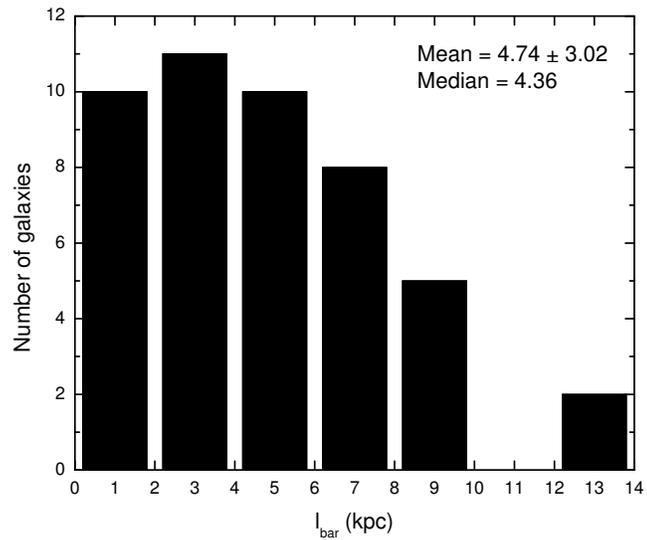}
\caption{Distribution of bar sizes for barred galaxies in our sample
(N=46).} \label{fig17}
\end{figure}

\begin{figure*}
\includegraphics[width=\columnwidth,clip=true]{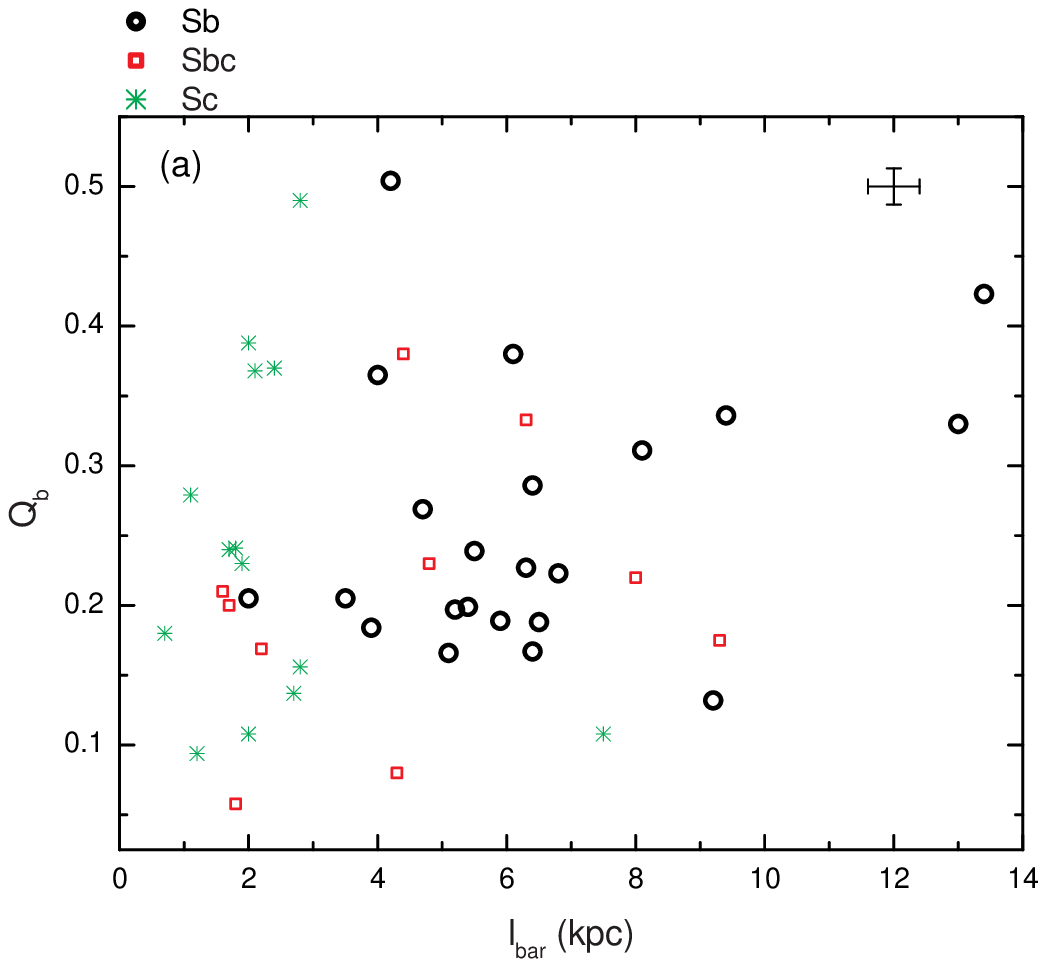}
\hspace{0.5cm}
\includegraphics[width=\columnwidth,clip=true]{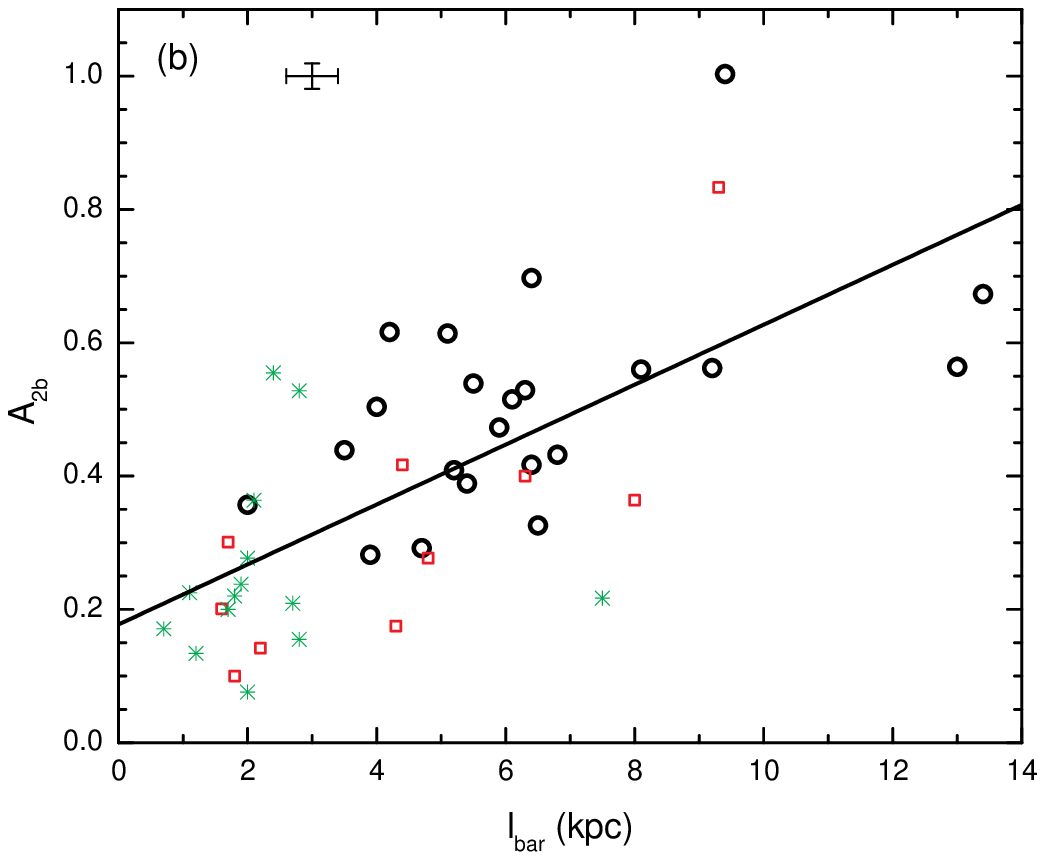}
\includegraphics[width=\columnwidth,clip=true]{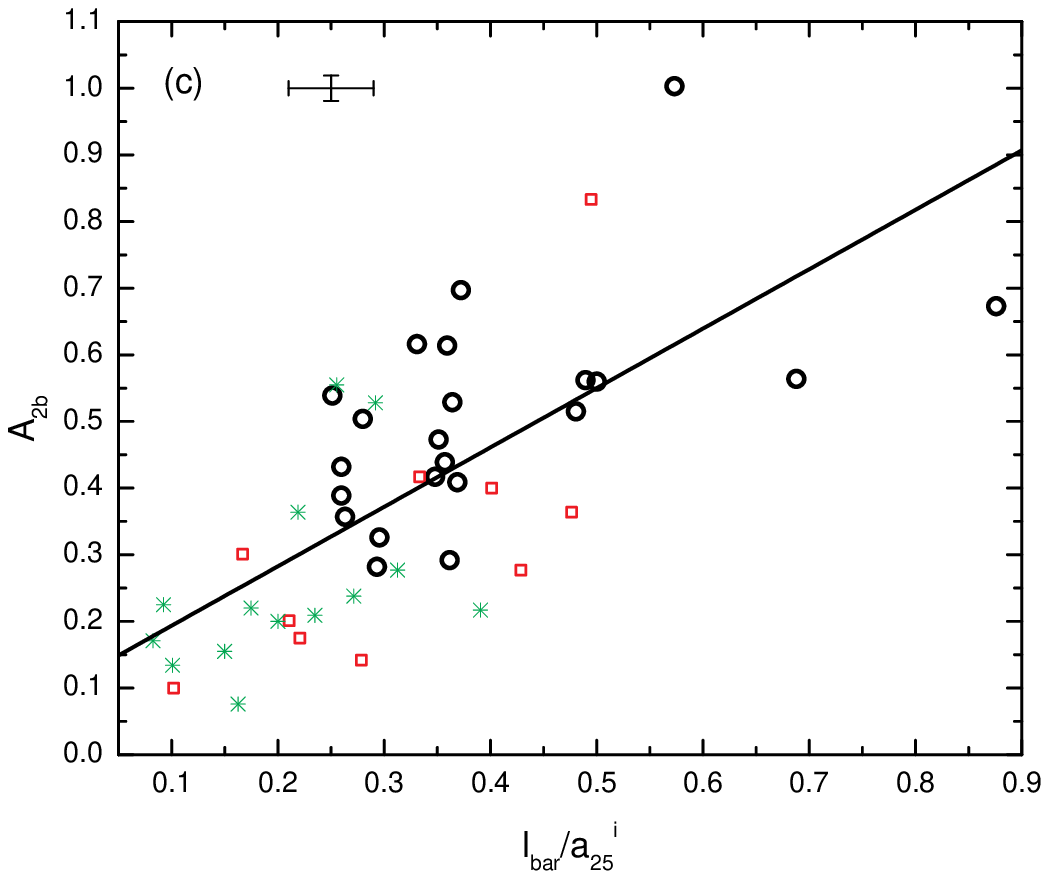}
\caption{Barred CIG/KIG Sb-Sc galaxies (N=46): (a) Bar strength
Q$_{b}$ versus maximum of the relative Fourier intensity amplitudes
at m=2, A$_{2b}$. (b) Maximum of the relative Fourier intensity
amplitudes at m=2, A$_{2b}$ versus bar size, l$_{bar}$. (c) A$_{2b}$
versus bar size, l$_{bar}$ normalized by the semimajor axis of the
25 mag arcsec$^{-2}$ isophote in i-band, a$_{25}^{i}$. Typical
2$\sigma$ error bars are shown in each panel. A linear regression is
shown in panels b and c.} \label{fig18}
\end{figure*}
\clearpage

\begin{figure*}
\includegraphics[width=0.92\columnwidth,clip=true]{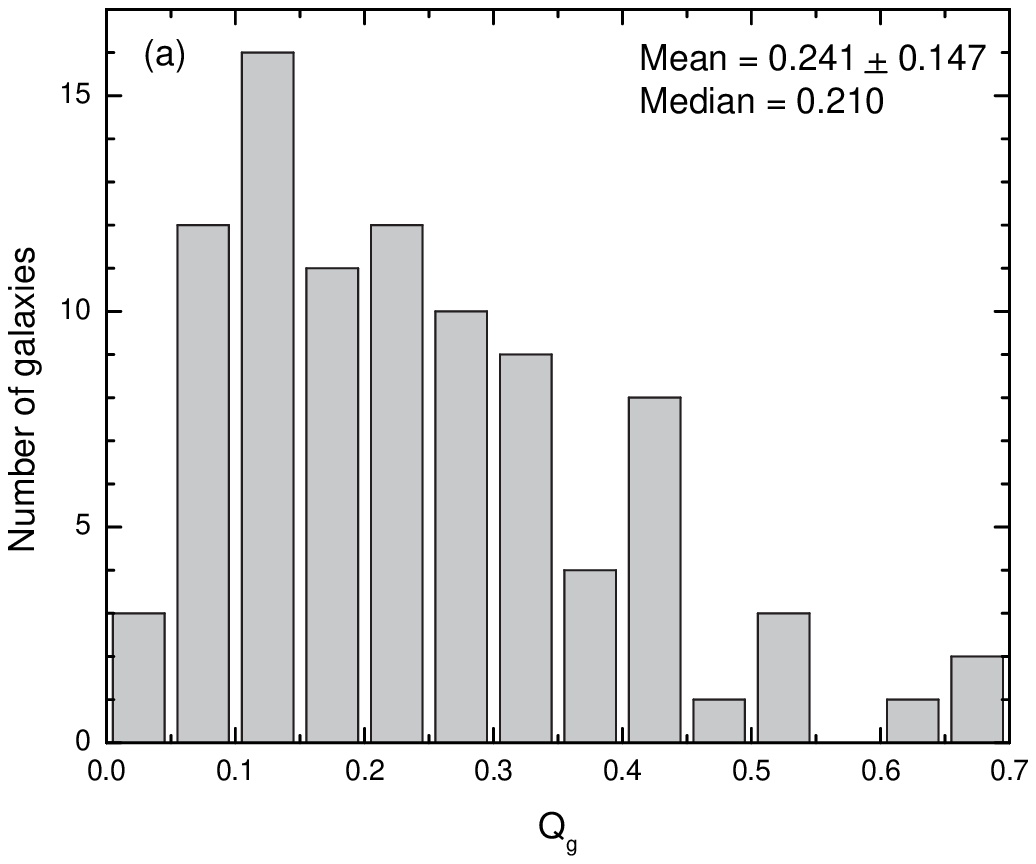}
\includegraphics[width=0.92\columnwidth,clip=true]{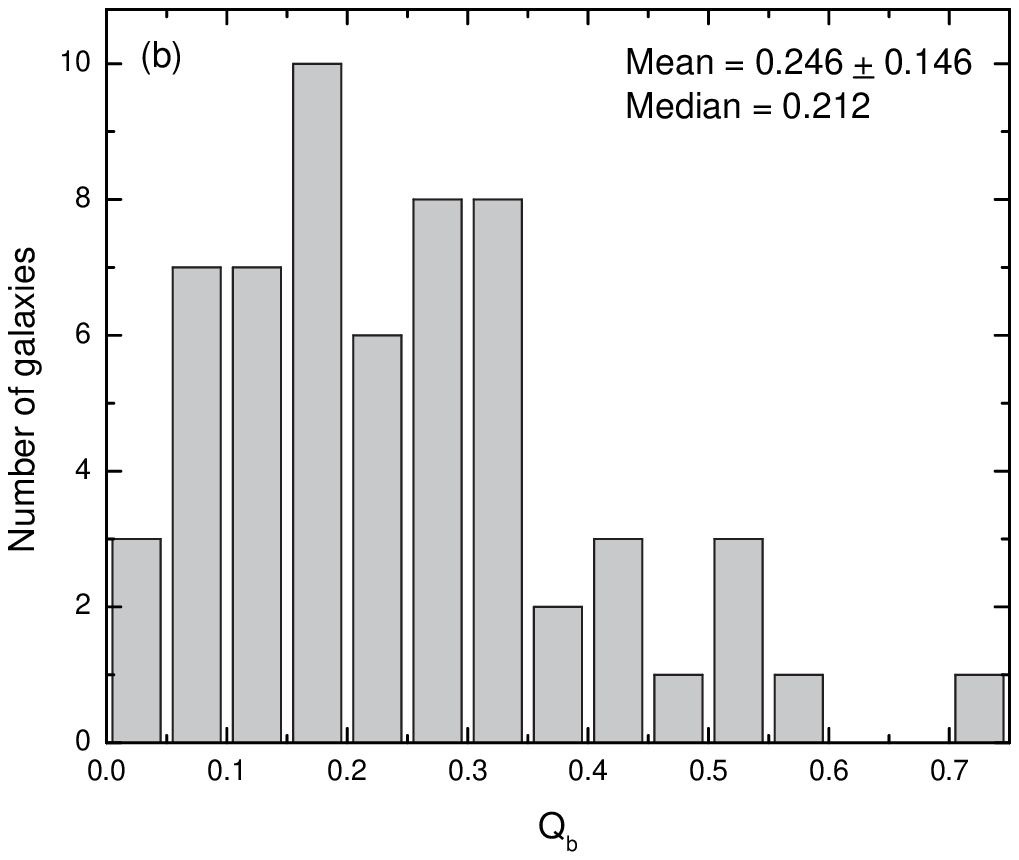}
\includegraphics[width=0.92\columnwidth,clip=true]{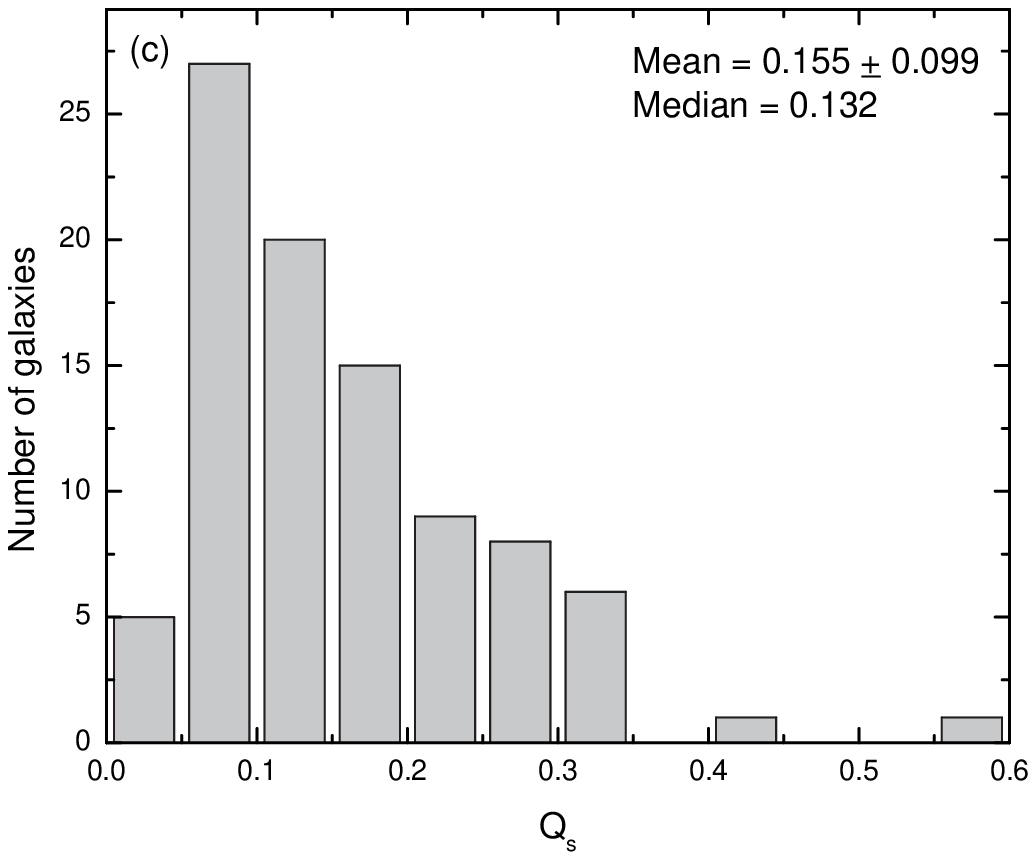}
\includegraphics[width=0.92\columnwidth,clip=true]{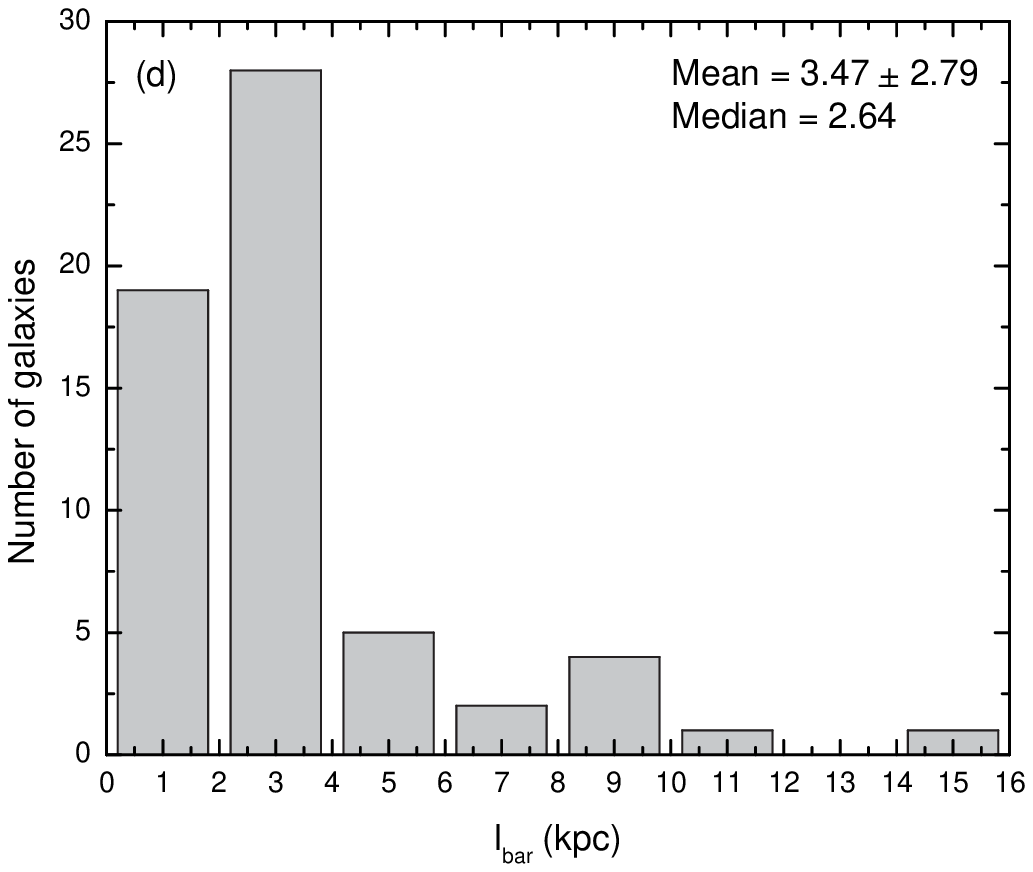}
\caption{(a) Distribution of the total strength Q$_{g}$ for the
Sb-Sc galaxies from OSU sample (N=92); (b) Distribution of the bar
strength Q$_{b}$ for the barred Sb-Sc galaxies from OSU sample
(N=60); (c) Distribution of the spiral strength Q$_{s}$ for the
Sb-Sc galaxies from OSU sample (N=92). (d) Distribution of bar sizes
for barred galaxies in the OSU sample (N=60).} \label{fig18}
\end{figure*}
\clearpage

\begin{figure}
\includegraphics[width=\columnwidth,clip=true]{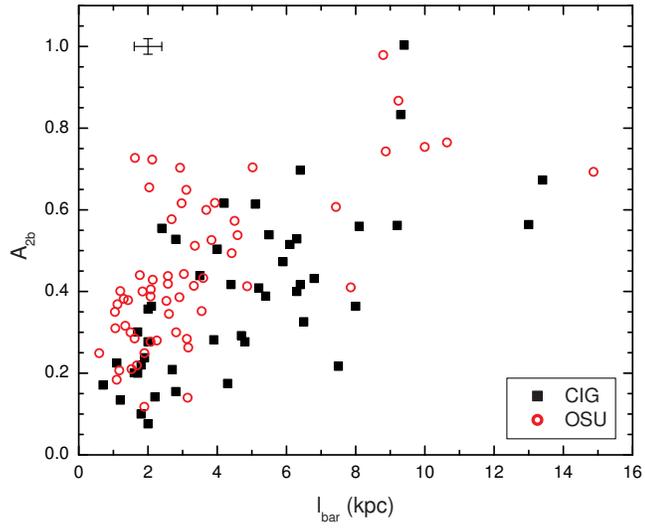}
\caption{A$_{2b}$ versus l$_{bar}$ for the barred galaxies in our
sample (N=46) and in the OSU sample (N=60). This shows that in
near-IR bands bars can be seen in higher contrast. Typical 2$\sigma$
error bars for the CIG galaxies are shown on the figure.}
\label{fig20}
\end{figure}
\clearpage

\begin{figure}
\includegraphics[width=\columnwidth,clip=true]{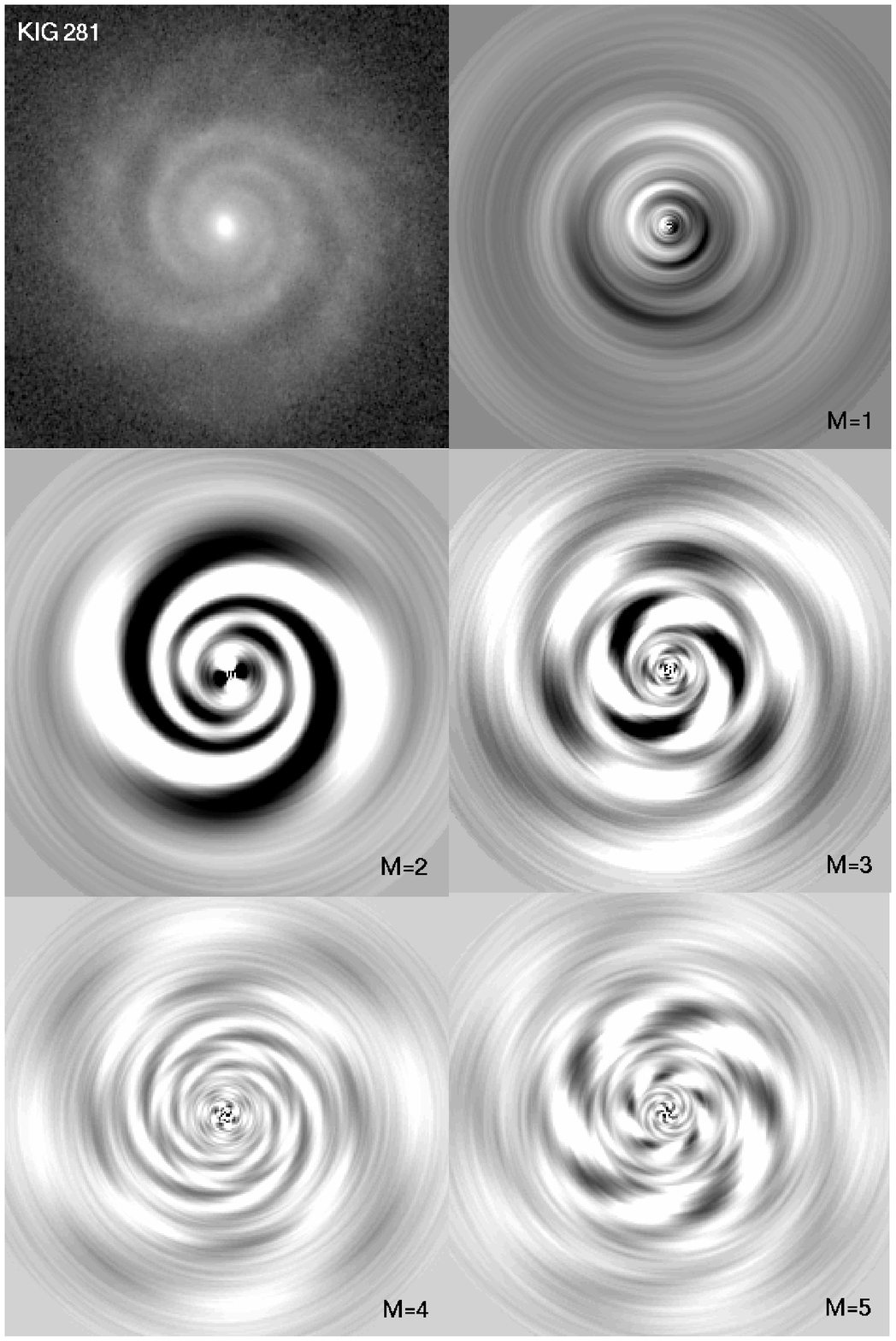}
\caption{KIG 281: The original reduced \& deprojected i-band image
and the reconstructed m = 1, 2, 3, 4, 5 Fourier term images.}
\label{fig21}
\end{figure}

\begin{figure}
\includegraphics[width=\columnwidth,clip=true]{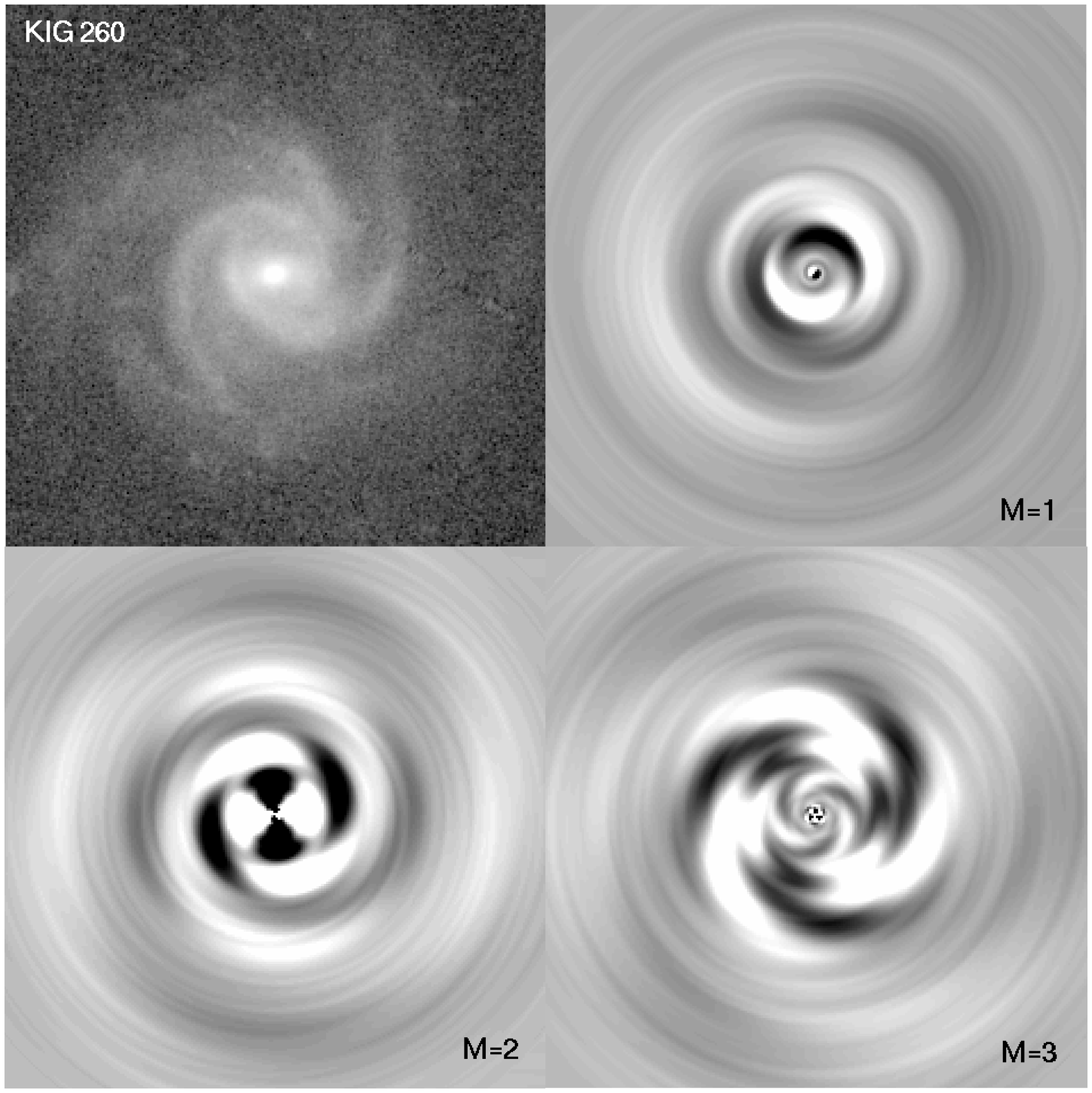}
\caption{KIG 260: The original reduced \& deprojected i-band image
and the reconstructed m = 1, 2, 3 Fourier term images.}
\label{fig22}
\end{figure}

\begin{figure}
\includegraphics[width=\columnwidth,clip=true]{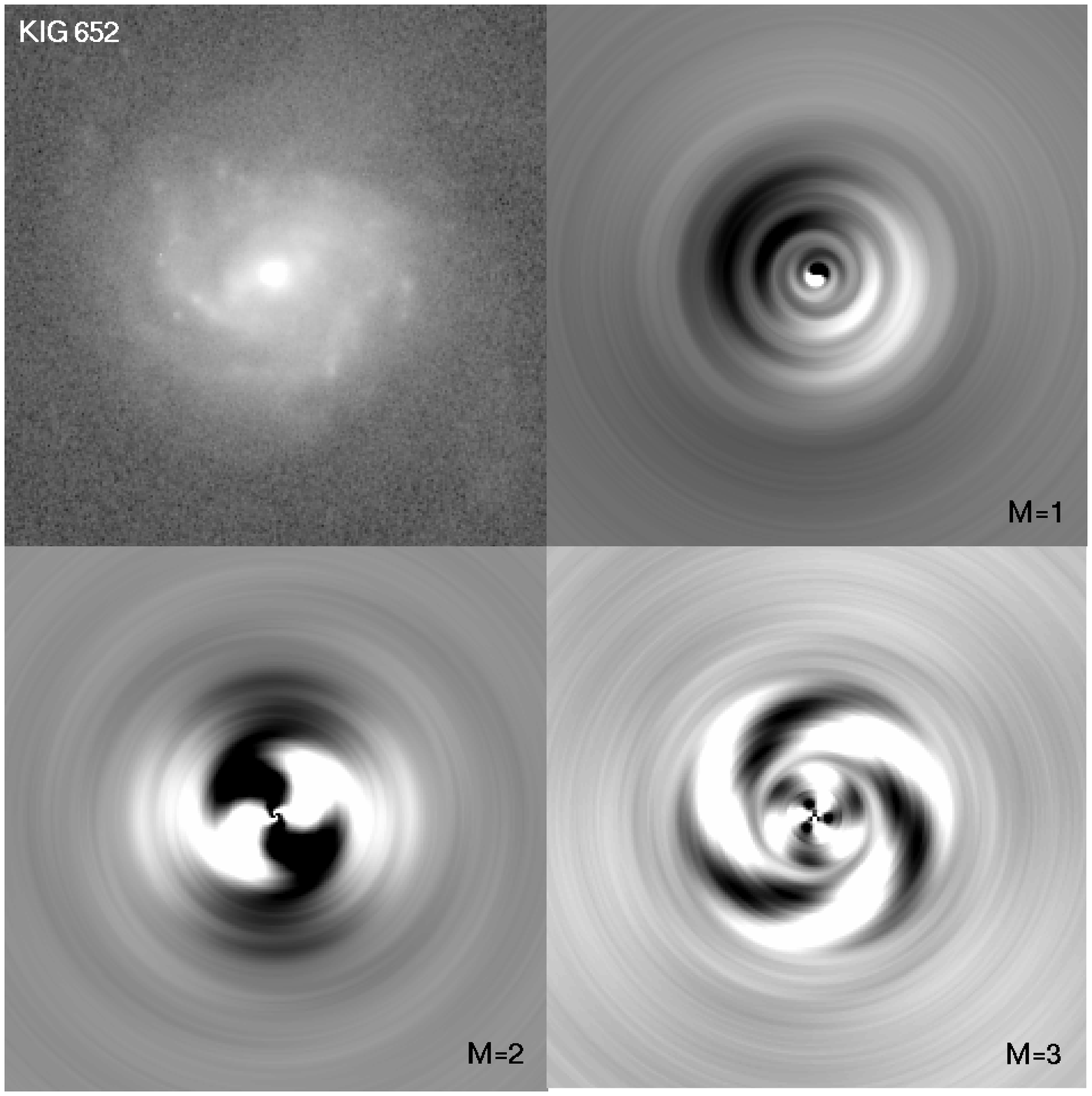}
\caption{KIG 652: The original reduced \& deprojected i-band image
and the reconstructed m = 1, 2, 3 Fourier term images.}
\label{fig23}
\end{figure}

\begin{figure}
\includegraphics[width=\columnwidth,clip=true]{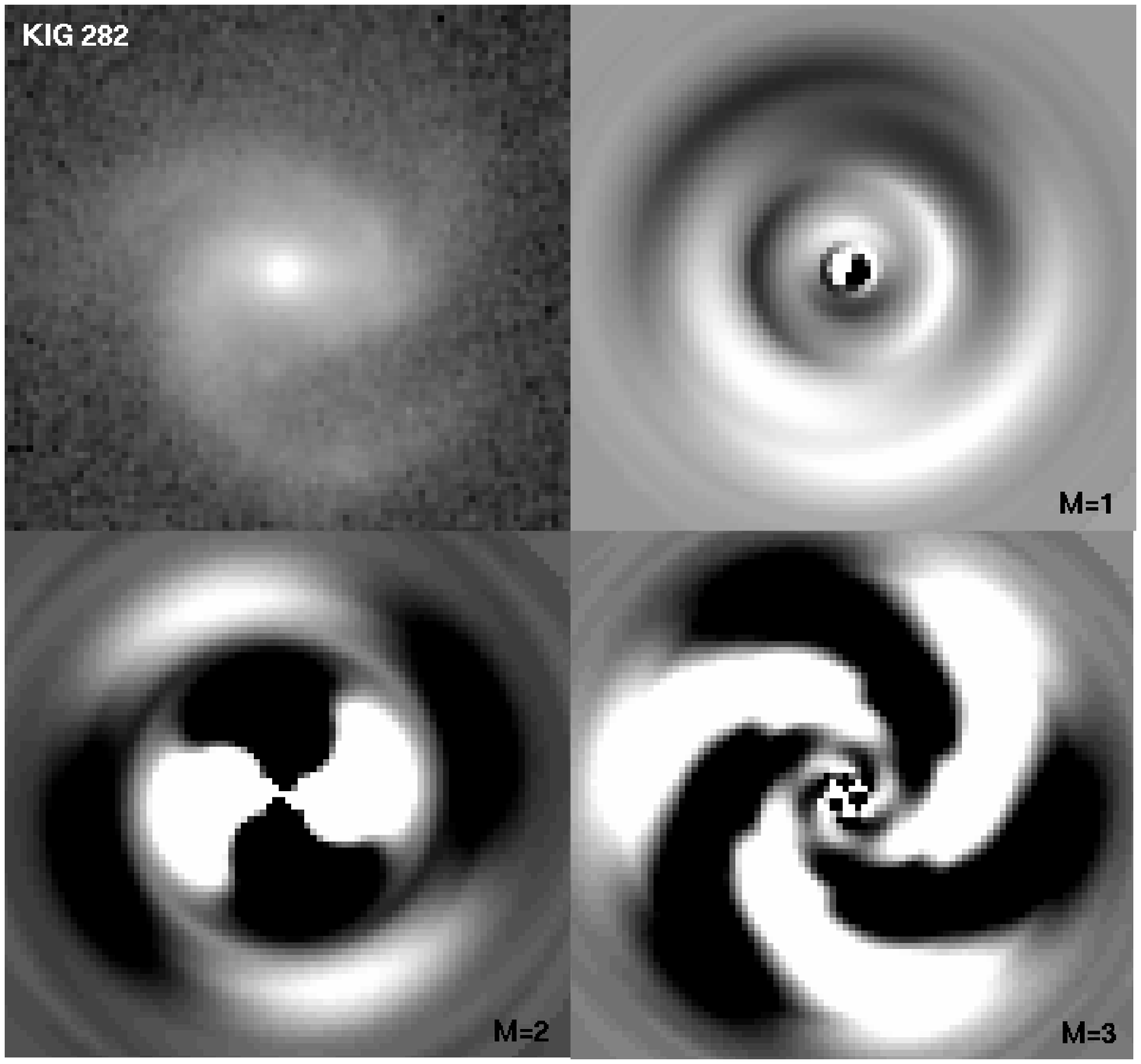}
\caption{KIG 282: The original reduced \& deprojected i-band image
and the reconstructed m = 1, 2, 3 Fourier term images.}
\label{fig24}
\end{figure}

\end{document}